\documentclass[twocolumn]{aastex61}
\submitjournal{ApJ}

\usepackage{hyperref}
\usepackage{amsmath}
\usepackage{amssymb}
\usepackage{graphicx}
\usepackage{enumitem}
\usepackage{lineno}

\newcommand{\Teff}{T_{\rm eff}}
\newcommand{\logg}{\log g}

\newcommand{\kms}{\text{km}\,\text{s}^{-1}}

\begin{document}
\title{Detailed Abundances in the Ultra-Faint Magellanic Satellites Carina II and III\footnote{This paper includes data gathered with the 6.5 meter Magellan Telescopes located at Las Campanas Observatory, Chile.}}
\shorttitle{Detailed Abundances in Car II and III}

\correspondingauthor{A.~P.~Ji}
\email{aji@carnegiescience.edu}

\author[0000-0002-4863-8842]{A.~P.~Ji}
\altaffiliation{Hubble Fellow}
\affiliation{Observatories of the Carnegie Institution for Science, 813 Santa Barbara St., Pasadena, CA 91101, USA}

\author[0000-0002-9110-6163]{T.~S.~Li}
\altaffiliation{NHFP Einstein Fellow}
\affiliation{Observatories of the Carnegie Institution for Science, 813 Santa Barbara St., Pasadena, CA 91101, USA}
\affiliation{Department of Astrophysical Sciences, Princeton University, Princeton, NJ 08544, USA}
\affiliation{Fermi National Accelerator Laboratory, P.O.\ Box 500, Batavia, IL 60510, USA}
\affiliation{Kavli Institute for Cosmological Physics, University of Chicago, Chicago, IL 60637, USA}

\author{J.~D.~Simon}
\affiliation{Observatories of the Carnegie Institution for Science, 813 Santa Barbara St., Pasadena, CA 91101, USA}
\author[0000-0003-0710-9474]{J.~Marshall}
\affiliation{George P. and Cynthia Woods Mitchell Institute for Fundamental Physics and Astronomy, and Department of Physics and Astronomy, Texas A\&M University, College Station, TX 77843, USA}
\author[0000-0003-4341-6172]{A.~K.~Vivas}
\affiliation{Cerro Tololo Inter-American Observatory, NSF's National Optical-Infrared Astronomy Research Laboratory, Casilla 603, La Serena, Chile}
\author[0000-0002-6021-8760]{A.~B.~Pace}
\altaffiliation{Mitchell Astronomy Fellow}
\affiliation{McWilliams Center for Cosmology, Carnegie Mellon University, 5000 Forbes Ave, Pittsburgh, PA 15213, USA}
\affiliation{George P. and Cynthia Woods Mitchell Institute for Fundamental Physics and Astronomy, and Department of Physics and Astronomy, Texas A\&M University, College Station, TX 77843, USA}
\author[0000-0001-8156-0429]{K.~Bechtol}
\affiliation{Physics Department, University of Wisconsin-Madison, 1150 University Avenue Madison, WI 53706, USA}
\author[0000-0001-8251-933X]{A.~Drlica-Wagner}
\affiliation{Fermi National Accelerator Laboratory, P.O.\ Box 500, Batavia, IL 60510, USA}
\affiliation{Kavli Institute for Cosmological Physics, University of Chicago, Chicago, IL 60637, USA}
\affiliation{Department of Astronomy and Astrophysics, University of Chicago, Chicago IL 60637, USA}
\author[0000-0003-2644-135X]{S.~E.~Koposov}
\affiliation{McWilliams Center for Cosmology, Carnegie Mellon University, 5000 Forbes Ave, Pittsburgh, PA 15213, USA}
\affiliation{Institute of Astronomy, University of Cambridge, Madingley Road, Cambridge CB3 0HA, UK}
\author[0000-0001-6154-8983]{T.~T.~Hansen}
\affiliation{George P. and Cynthia Woods Mitchell Institute for Fundamental Physics and Astronomy, and Department of Physics and Astronomy, Texas A\&M University, College Station, TX 77843, USA}

\author{S.~Allam}
\affiliation{Fermi National Accelerator Laboratory, P.O. Box 500, Batavia, IL 60510, USA}
\author{R.~A.~Gruendl}
\affiliation{Department of Astronomy, University of Illinois, 1002 W. Green Street, Urbana, IL 61801, USA}
\affiliation{National Center for Supercomputing Applications, 1205 West Clark St., Urbana, IL 61801, USA}
\author{M.~D.~Johnson}
\affiliation{National Center for Supercomputing Applications, 1205 West Clark St., Urbana, IL 61801, USA}
\author{M.~McNanna}
\affiliation{Physics Department, University of Wisconsin-Madison, 1150 University Avenue Madison, WI 53706, USA}
\author{N.~E.~D.~No\"el}
\affiliation{Department of Physics, University of Surrey, Guildford, GU2 7XH, UK}
\author{D.~L.~Tucker}
\affiliation{Fermi National Accelerator Laboratory, P.O. Box 500, Batavia, IL 60510, USA}
\author{A.~R.~Walker}
\affiliation{Cerro Tololo Inter-American Observatory, NSF's National Optical-Infrared Astronomy Research Laboratory, Casilla 603, La Serena, Chile}

\collaboration{(MagLiteS Collaboration)}

\begin{abstract}
We present the first detailed elemental abundances in the ultra-faint Magellanic satellite galaxies Carina~II (Car~II) and Carina~III (Car~III).
With high-resolution Magellan/MIKE spectroscopy, we determined abundances of nine stars in Car~II including the first abundances of an RR Lyrae star in an ultra-faint dwarf galaxy; and two stars in Car~III.
The chemical abundances demonstrate that both systems are clearly galaxies and not globular clusters.
The stars in these galaxies mostly display abundance trends matching those of other similarly faint dwarf galaxies:
enhanced but declining [$\alpha$/Fe] ratios, iron-peak elements matching the stellar halo, and unusually low neutron-capture element abundances. One star displays a low outlying [Sc/Fe]$=-1.0$.
We detect a large Ba scatter in Car~II, likely due to inhomogeneous enrichment by low-mass AGB star winds.
The most striking abundance trend is for [Mg/Ca] in Car~II, which decreases from $+0.4$ to $-0.4$ and indicates clear variation in the initial progenitor masses of enriching core-collapse supernovae.
So far, the only ultra-faint dwarf galaxies displaying a similar [Mg/Ca] trend are likely satellites of the Large Magellanic Cloud.
We find two stars with $\mbox{[Fe/H]} \leq -3.5$, whose abundances likely trace the first generation of metal-free Population~III stars and are well-fit by Population~III core-collapse supernova yields.
An appendix describes our new abundance uncertainty analysis that propagates line-by-line stellar parameter uncertainties.
\end{abstract}
\keywords{stars: abundances --- galaxies: dwarf --- Local Group}

\section{Introduction}

Ultra-faint dwarf galaxies (UFDs) are the luminous counterparts to the least massive star-forming dark matter halos, likely forming stars during the first $\sim$1 Gyr before being quenched by reionization \citep[e.g.,][]{Bullock00,Benson02,Simon07,Brown14,Simon19}.
As a result, the chemical abundances of stars in UFDs preserve a clean snapshot of chemical enrichment from the earliest stages of galaxy formation and reionization, providing a window to the most metal-poor stellar populations and their nucleosynthetic output \citep{Kirby08,Frebel12,Geha13,Weisz14c,Wise14,Ji15}.
Dozens of UFDs have now been discovered in deep, wide, and uniform photometric surveys such as the Sloan Digital Sky Survey, Pan-STARRS, and the Dark Energy Survey (DES) \citep[e.g.,][]{Willman05a,Belokurov07,Laevens15a,Bechtol15,Koposov15a,DrlicaWagner15b}.
The large number of UFDs provides a large population of local objects that retain signatures of high-redshift star and galaxy formation.

Until recently, these UFDs have generally been assumed to be satellites of the Milky Way.  However, the two most massive dwarfs orbiting the Milky Way, the Large and Small Magellanic Clouds (LMC and SMC), should have had their own satellite UFDs \citep[e.g.,][]{DOnghia08,Koposov15b,DrlicaWagner15b,Jethwa16,Dooley17,Sales17}.  Since the LMC and SMC are likely on their first infall into the Milky Way \citep{Besla07,Busha11,Kallivayalil13,Simon18,Fritz19,Pace19}, any dwarfs that were previously Magellanic satellites could now be in the process of accretion into the Milky Way.  Gaia proper motion measurements have revealed that several UFDs are kinematically associated with the LMC/SMC system \citep{Kallivayalil18,Erkal19}.
Two of these LMC satellites are Carina~II (Car~II, $M_V=-4.5$, $L/L_\odot \sim 10^{3.7}$) and Carina~III (Car~III, $M_V=-2.4$, $L/L_\odot \sim 10^{2.9}$), discovered in the Magellanic Satellites Survey (MagLiteS, \citealt{DrlicaWagner16,Torrealba18}) with the Dark Energy Camera (DECam, \citealt{Flaugher15}) on the Blanco telescope.
\citet{Li18Carina} spectroscopically confirmed Car~II to be a dwarf galaxy, and \citet{LiPrep} have now confirmed Car~III as a dwarf galaxy as well.
These UFDs are only ${\sim}20$ kpc away from the LMC, and are also close to the Sun (37.4 and 27.8 kpc for Car~II and III, respectively). Thus, they have a relatively large number of bright stars amenable for high-resolution spectroscopic followup and chemical abundance measurements.

In this paper, we present a comprehensive chemical abundance analysis of Magellan/MIKE spectroscopy of 9 stars in Car~II and 2 stars in Car~III.
Along with Horologium~I \citep{Nagasawa18}, these are currently the only ultra-faint LMC satellites with high-resolution abundance measurements.
Section~\ref{sec:observations} explains the observations, data reduction, and velocity measurements.
Section~\ref{sec:abundances} details our abundance analysis.
We discuss the formation history of these galaxies in Section~\ref{sec:formation}, highlighting the interesting $\alpha$-element abundance trends in Section~\ref{sec:alphaevol}.
We focus on potential signatures of metal-free Pop~III stars in Section~\ref{sec:pop3},
then summarize and conclude in Section~\ref{sec:conclusion}.

\section{Observations, Data Reduction, Radial Velocities}\label{sec:observations}
\begin{deluxetable*}{lcccccccccc}
\tablecolumns{11}
\tabletypesize{\footnotesize}
\tablecaption{\label{tab:obs}Observations}
\tablehead{Star & \texttt{source\_id} & RA & Dec & $G$ & $g_0$ & $r_0$ & Slit & $t_{\rm exp}$ & SNR & SNR \\
{}&{}&{}&{}&(mag)&(mag)&(mag)&{}&(hour)&(4500{\AA})&(6500{\AA})}
\startdata
CarII-6544  & 5293947247051916544 & 07:36:51.11 & $-$58:01:46.3 & 15.07 & 15.63 & 14.63 & 0\farcs5 & 1.8 & 22 & 67 \\
CarII-7872  & 5293894539213647872 & 07:36:51.89 & $-$58:16:39.2 & 15.50 & 15.92 & 15.01 & 0\farcs7 & 1.0 & 25 & 60 \\
CarII-5664  & 5293896360279425664 & 07:38:08.51 & $-$58:09:35.0 & 16.33 & 16.55 & 15.86 & 0\farcs7 & 3.8 & 38 & 80 \\
CarII-0064  & 5293951473299720064 & 07:36:21.25 & $-$57:58:00.2 & 16.78 & 16.96 & 16.30 & 1\farcs0 & 2.6 & 22 & 54 \\
CarII-4704  & 5293928074318184704 & 07:35:37.66 & $-$58:01:51.8 & 17.40 & 17.46 & 16.93 & 0\farcs7 & 3.3 & 13 & 34 \\
CarII-9296  & 5293900827045399296 & 07:37:39.79 & $-$58:05:06.9 & 17.72 & 17.86 & 17.29 & 1\farcs0 & 3.0 & 15 & 35 \\
CarII-2064  & 5293951881319592064 & 07:36:01.33 & $-$57:58:43.8 & 18.22 & 18.27 & 17.77 & 0\farcs7 & 4.6 & 13 & 31 \\
CarII-4928  & 5293951503362524928 & 07:36:24.98 & $-$57:57:14.2 & 18.42 & 18.40 & 17.96 & 1\farcs0 & 5.5 & 13 & 31 \\
CarII-V3\tablenotemark{*}  & 5293940924860019584 & 07:35:09.12 & $-$57:57:14.8 & 18.46 & 18.13 & 18.01 & 1\farcs0 & 2.5 & 15 & 24 \\
CarIII-1120 & 5293955665187701120 & 07:38:22.30 & $-$57:53:02.1 & 17.46 & 17.51 & 16.97 & 0\farcs7 & 2.7 & 18 & 39 \\
CarIII-8144 & 5293907630273478144 & 07:38:34.93 & $-$57:57:05.3 & 17.65 & 17.72 & 17.18 & 0\farcs7 & 3.2 & 21 & 41 \\
\enddata
\tablecomments{
Our star ID numbers are the last four digits of the Gaia \texttt{source\_id}.
$G$ is Gaia magnitudes. $g_0$ and $r_0$ are dereddened DECam photometry from MagLiteS, taken from \citet{LiPrep}. SNR is per pixel
}
\tablenotetext{*}{This star is a variable RR Lyrae star. The magnitudes here are the mean magnitudes found by MagLiteS and \emph{Gaia} \citep{Torrealba18,Clementini19}, where the DECam magnitudes have been dereddened.}
\end{deluxetable*}

\begin{deluxetable}{lccccc}
\tablecolumns{6}
\tabletypesize{\footnotesize}
\tablecaption{\label{tab:rv}Radial Velocities}
\tablehead{Star & Obs Date & MJD & $v_{\text{hel}}$ & $N_{\text{ord}}$ & $\sigma_{\text{sys}}$}
\startdata
CarII-6544  & 2018-01-24 & 58142.031 & 470.4 & 34 & 1.2 \\
CarII-7872  & 2018-11-13 & 58435.277 & 478.5 & 35 & 0.5 \\
CarII-5664  & 2018-11-16 & 58438.207 & 483.7 & 34 & 0.6 \\
CarII-0064  & 2017-12-06 & 58093.336 & 475.0 & 35 & 0.7 \\
CarII-4704  & 2018-11-13 & 58435.319 & 472.4 & 30 & 1.0 \\
CarII-9296  & 2018-01-24 & 58142.059 & 481.9 & 30 & 1.2 \\
CarII-2064  & 2018-01-24 & 58142.210 & 473.9 & 34 & 0.9 \\
CarII-4928  & 2018-01-23 & 58141.220 & 476.1 & 23 & 1.4 \\
CarII-V3    & 2018-11-15 & 58437.247 & 478.2 & 18 & 1.9 \\
CarIII-1120 & 2018-01-24 & 58142.147 & 283.7 & 30 & 1.1 \\
CarIII-8144 & 2018-11-16 & 58438.277 & 280.8 & 36 & 0.5 \\
\enddata
\tablecomments{We show one representative velocity measurement per star in our sample. The full table is available online. Note that CarII-6544 is likely a binary star and CarII-V3 is an RRL star, so these have significant velocity variations.}
\end{deluxetable}

Our Carina~II and III targets were selected to be the brightest radial velocity members from Magellan/IMACS, AAT/AAO, and VLT/FLAMES moderate resolution spectra, including five bright member stars from \citet{Li18Carina} and five new bright member stars from \citet{LiPrep}. In addition, we include one RR Lyrae member in Carina~II identified in \citet{Torrealba18}.
We observed these stars with Magellan/MIKE \citep{Bernstein03} over four separate runs (Tables~\ref{tab:obs} and \ref{tab:rv}).
Slits of width 0\farcs5, 0\farcs7, and 1\farcs0 were used depending on the seeing, resulting in typical resolutions of $R \sim $50k/40k, 35k/28k, and 28k/22k on the blue/red arms of MIKE, respectively.
We used 2x2 binning for the 0\farcs7 and 1\farcs0 slits, and 2x1 binning for the 0\farcs5 slit.
The MIKE data were reduced with CarPy \citep{Kelson03}.

\begin{figure*}
    \centering
    \includegraphics[width=0.9\linewidth]{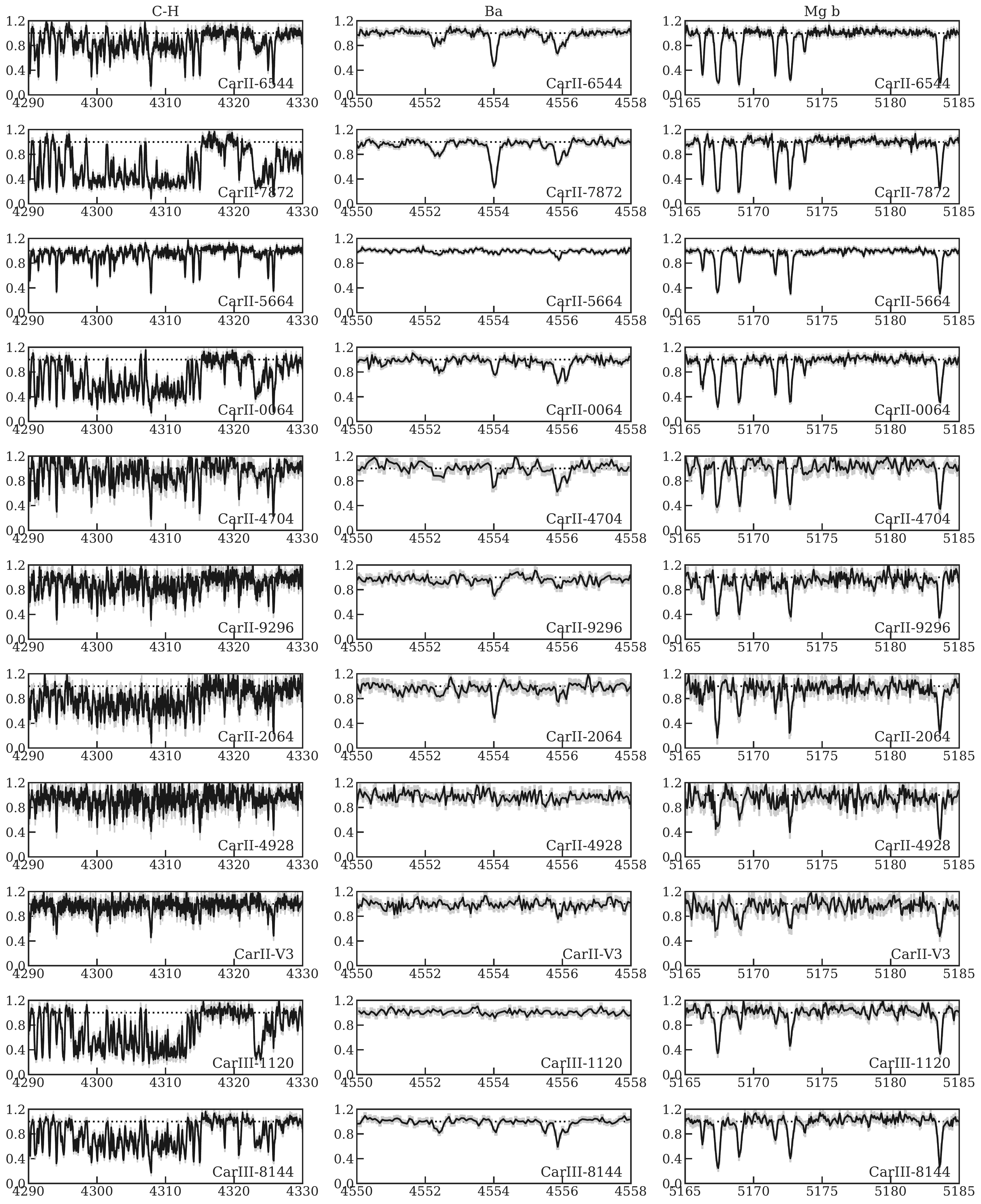}
    \caption{Spectrum of all stars around the C-H G band, the Ba 4554 line, and the Mg b lines. Stars are sorted in order of system and then increasing $\Teff$ from top to bottom (same as Table~\ref{tab:sp}). Grey band indicates ${\pm}1\sigma$ spectrum noise.
    }
    \label{fig:specs}
\end{figure*}

We used the code SMHR \citep{Casey14}\footnote{\url{https://github.com/andycasey/smhr}, first described in \citealt{Casey14}} to coadd, normalize, stitch orders, and Doppler correct the reduced spectra for abundance analysis.
Data from multiple runs were combined by coadding order-by-order, using a common set of spline knot locations and line masks after adjusting for observed radial velocity.
The signal-to-noise at the order center closest to rest wavelengths of 4500{\AA}, 5300{\AA}, and 6500{\AA} is given in Table~\ref{tab:obs}.
The total integrated time spent on these stars is 34 hours.
Note there is significant reddening towards Car~II and III (E($B-V$) $\sim 0.2$ mag).
Figure~\ref{fig:specs} shows our spectra around the C-H G band, the strongest barium line, and the Mg b triplet.

In general, we reduced all MIKE data from a given observing run together before measuring the radial velocity.
The exception is the RR Lyrae (RRL) star CarII-V3, which experiences large radial velocity variations on a short timescale.
Using the known pulsation phases \citep{Torrealba18},
we observed CarII-V3 across phases $0.40-0.55$ with five consecutive 30-min exposures.
Over this phase range, the star has fairly consistent stellar parameters \citep{For11}, so individual exposures can be coadded after correcting for a velocity offset.
We reduced each exposure separately, measured radial velocities for each observation separately using the Mg b triplet, corrected each order to rest frame, and coadded order-by-order before stitching orders in SMHR.

Radial velocities are given in Table~\ref{tab:rv}. For the velocity measurements, we re-reduced each exposure individually with CarPy. We measured radial velocities of the 40 orders from 3900\AA\ to 6800\AA\ (order numbers $51-90$). Of these, we masked the telluric lines around 6300\AA, discarded three orders from 5820-6020\AA\ because of interstellar Na D absorption, and discarded the bluest order on the red side due to uniformly low S/N. We cross-correlated individual orders of our MIKE spectra against a normalized high-S/N MIKE spectrum of HD122563. To remove outliers, we iteratively sigma clip orders with velocities that are more than 5 biweight scales away from the biweight average.
The final number of orders for each spectrum is given by $N_{\text{ord}}$ in Table~\ref{tab:rv}.
Statistical errors for each order were then found by calculating the $\chi^2$ at different velocities and taking $\Delta \chi^2=1$ away from the minimum.

Naively, we could combine these measurements by taking a weighted average of all orders to get a final average velocity and in principle reaching an extremely high velocity precision of ${\sim}0.1$ km/s.
However, systematic effects dominate both the velocity measurement and error.
For example, MIKE is not attached to the instrument rotator and until recently did not have an atmospheric dispersion compensator. At high airmasses, atmospheric refraction in the narrow slit direction causes systematic velocity offsets as a function of wavelength that can be as large as $2-3$ km/s.
We will correct for these effects in later work, but such velocity differences do not impact the abundance analyses that are the focus of this paper.
Thus, for now in Table~\ref{tab:rv} we provide the radial velocity of each individual spectrum computed by an inverse-variance weighted average of all $N_{\text{ord}}$ orders. The systematic error is the weighted standard deviation of those orders and dominates over the ${\sim}0.1$ km/s statistical uncertainty.

\section{Abundance Analysis}\label{sec:abundances}
\begin{deluxetable}{lcccc}
\tablecolumns{5}
\tabletypesize{\footnotesize}
\tablecaption{\label{tab:sp}Stellar Parameters}
\tablehead{Star & $\Teff$ (K) & $\logg$ (dex) & $\nu_t$ ($\kms$) & [M/H]}
\startdata
CarII-6544  & $4330 \pm 152$ & $0.40 \pm 0.31$ & $2.75 \pm 0.26$ & $-2.65 \pm 0.09$ \\
CarII-7872  & $4380 \pm 155$ & $0.75 \pm 0.32$ & $2.32 \pm 0.27$ & $-2.48 \pm 0.11$ \\
CarII-5664  & $4430 \pm 155$ & $0.45 \pm 0.31$ & $2.34 \pm 0.25$ & $-3.50 \pm 0.06$ \\
CarII-0064  & $4630 \pm 153$ & $1.15 \pm 0.32$ & $2.31 \pm 0.27$ & $-2.20 \pm 0.07$ \\
CarII-4704  & $4720 \pm 160$ & $1.30 \pm 0.31$ & $1.97 \pm 0.27$ & $-2.19 \pm 0.09$ \\
CarII-9296  & $4810 \pm 205$ & $1.40 \pm 0.37$ & $1.90 \pm 0.34$ & $-2.87 \pm 0.15$ \\
CarII-2064  & $5300 \pm 200$ & $2.70 \pm 0.35$ & $2.15 \pm 0.32$ & $-2.35 \pm 0.17$ \\
CarII-4928  & $5065 \pm 236$ & $2.35 \pm 0.46$ & $2.10 \pm 0.34$ & $-3.00 \pm 0.20$ \\
CarII-V3    & $6100 \pm 330$ & $1.75 \pm 0.27$ & $3.20 \pm 0.28$ & $-2.70 \pm 0.21$ \\
CarIII-1120 & $4500 \pm 216$ & $1.50 \pm 0.34$ & $1.85 \pm 0.32$ & $-3.89 \pm 0.14$ \\
CarIII-8144 & $4990 \pm 162$ & $2.20 \pm 0.32$ & $1.75 \pm 0.27$ & $-2.25 \pm 0.08$ \\
\enddata
\vspace{-3em}
\end{deluxetable}

\begin{deluxetable*}{lrrrrrrrrrrrrrrrrr}
\tablecolumns{17}
\tabletypesize{\footnotesize}
\tablecaption{\label{tab:lines}Line Measurements}
\tablehead{Star & $\lambda$ & ID & $\chi$ & $\log gf$ & EW & $\sigma$(EW) & ul & $\log \epsilon$ & $\sigma_i$ & $\sigma_{i,\text{stat}}$ & $\sigma_{i,\text{sys}}$ & $\delta_{i,\Teff}$ & $\delta_{i,\logg}$ & $\delta_{i,u_t}$ & $\delta_{i,\text{[M/H]}}$ & $\sigma_{\text{cont}}$}
\startdata
CarII-0064  & 5183.60 &  12.0 &  2.72 & -0.17 & 251.8 &  7.0 & 0 &  5.21 &  0.28 &  0.04 &  0.28 & +0.23 & -0.14 & -0.08 & -0.01 &  0.07  \\
CarII-0064  & 5528.40 &  12.0 &  4.35 & -0.50 &  87.0 &  4.7 & 0 &  5.35 &  0.14 &  0.07 &  0.12 & +0.10 & -0.05 & -0.05 & -0.01 &  0.04  \\
CarII-0064  & 4323.00 & 106.0 &\nodata&\nodata&  syn  &  syn & 0 &  6.16 &  0.30 &  0.03 &  0.30 & +0.28 & -0.07 & +0.01 & +0.03 &\nodata \\
CarII-0064  & 4554.00 &  56.1 &  0.00 &  0.16 &  syn  &  syn & 0 & -2.52 &  0.17 &  0.08 &  0.15 & +0.10 & +0.11 & -0.01 & +0.01 &  0.03  \\
CarII-0064  & 4129.70 &  63.1 &  0.00 &  0.22 &  syn  &  syn & 1 & -1.89 &\nodata&\nodata&\nodata&\nodata&\nodata&\nodata&\nodata&\nodata \\
\enddata
\tablecomments{A portion of this table is shown for form. The full version is available online. See Appendix~\ref{app:error} for column details.}
\end{deluxetable*}

\subsection{Abundance Analysis Details}
We performed a standard 1D-LTE analysis using the 2017 version of the 1D LTE radiative transfer code MOOG \citep{Sneden73, Sobeck11}\footnote{\url{https://github.com/alexji/moog17scat}} and the \citet{Castelli04} (ATLAS) model atmospheres.
We used SMHR to measure equivalent widths, interpolate model atmospheres, and run MOOG.

For the red giant branch (RGB) stars, stellar parameters were derived spectroscopically. Briefly, we start assuming $\alpha$-enhanced $[\alpha/\text{Fe}]=+0.4$ model atmospheres. The effective temperature, surface gravity, and microturbulence ($\Teff$, $\logg$, $\nu_t$) were determined by balancing excitation, ionization, and line strength for Fe lines, respectively.
We then applied the $\Teff$ correction from \citet{Frebel13} to place the measurements on a photometric temperature scale and redetermined $\logg$ and $\nu_t$.
After this initial determination, if the star turned out to have low Mg abundances, we switched to $[\alpha/\text{Fe}]=0$ atmospheres and redetermined the stellar parameters.
Statistical stellar parameter uncertainties are found following \citet{Ji19a}, and we adopt systematic uncertainties of 150 K for $\Teff$, 0.3 dex for $\logg$, and $0.2\,\kms$ for $\nu_t$ due to uncertainties in the \citet{Frebel13} temperature calibration.
The statistical and systematic uncertainties were added in quadrature to obtain the total stellar parameter uncertainties in Table~\ref{tab:sp}.

We used a combination of equivalent widths and spectral syntheses to measure the abundances of individual lines.
We also determined statistical and systematic abundance uncertainties \emph{for each individual feature}.
For lines measured using equivalent widths, we propagated the $1\sigma$ equivalent width uncertainty into a $1\sigma$ statistical abundance uncertainty.
For lines measured using syntheses, we increased the element abundance until $\Delta\chi^2=1$, also corresponding to a $1\sigma$ statistical uncertainty.
These uncertainties account for continuum placement uncertainty (see Appendix~\ref{app:error} for details).
For the systematic uncertainties, we varied each stellar parameter ($\Teff$, $\logg$, $\nu_t$, [M/H]) individually by its error and remeasured the abundance.
The total systematic uncertainty is the quadrature sum of the individual stellar parameter uncertainties.
Finally, the total abundance uncertainty \emph{for an individual line} is the quadrature sum of the statistical and systematic uncertainty.
Individual line measurements and uncertainties are found in Table~\ref{tab:lines}.

We use inverse-variance weighted averages to combine lines into a final abundance.
Because we have included a detailed account of line-by-line uncertainties, this automatically downweights lines in regions of low spectral S/N; saturated lines that are sensitive to small equivalent width variations; and lines that are particularly sensitive to stellar parameters.
We verified that the weighted averages are usually only a few hundredths of a dex different from the unweighted averages. The exception is elements with few measurable lines like Si and Al, where some lines are much lower quality than others.
See Appendix~\ref{app:error} for detailed equations.

[X/Fe] ratios are derived by taking ratios of common ionization states (e.g., [Mg\,I/Fe\,I], [Ti\,II/Fe\,II]).
This mostly (though not always) results in smaller [X/Fe] errors than [X/H] errors, since some stellar parameter differences cancel out.
We also consistently propagate stellar parameter uncertainties for [X/Y] ratios, such as [Mg/Ca].

Upper limits were derived by spectrum synthesis. For a given feature, we fit a synthetic spectrum that well-matched the observed spectrum to determine a reference $\chi^2$ and local spectrum smoothing.
Then holding the continuum and smoothing fixed, we increased the abundance until $\Delta\chi^2 = 25$. This is formally a $5\sigma$ upper limit but does not include uncertainties in continuum placement.

\subsection{Abundance corrections}
Various systematics can affect 1D-LTE abundances of red giants. We tabulate several abundance corrections in Table~\ref{tab:corr}, which are the average of line-by-line corrections. These corrections have been applied in all figures but \emph{not} in Tables~\ref{tab:lines} or \ref{tab:abunds}.

Carbon is systematically converted to nitrogen in evolved red giants due to CN cycling. We estimate the natal carbon abundances of these stars with the corrections from \citet{Placco14}\footnote{\url{http://vplacco.pythonanywhere.com/}}.
Hotter stars have no correction, while for cooler/more evolved stars the correction can be as large as $+0.75$ dex.
We use the default correction grid assuming [N/Fe]$=0$, but changing [N/Fe] makes minimal difference.
Note that we assume all our stars are on the RGB, but if we had red clump or AGB stars in our sample they would have larger carbon corrections than applied here.

Only the Na D lines are available for sodium abundances, and these can have fairly large negative NLTE corrections.
We apply Na corrections from \citet{Lind11}\footnote{\url{www.inspect-stars.com}}, which range from $-0.13$ to $-0.48$ dex.
For CarII-6544 and CarII-7872, and CarII-5664 we set $\logg = 1$ to avoid the edge of the corrections grid.

Mg is marginally affected by NLTE effects in our stars. However, since Mg will be a very important element later, we tabulate the NLTE corrections just to show they are only affected by $<0.04$ dex \citep{Osorio15a,Osorio15b}.
For several stars (CarII-6544, CarII-4704, CarII-0064, CarII-5664, CarII-7872) we set $\logg = 1.5$ to avoid the edge of the corrections grid.
Note that we have used the two high-equivalent width Mg b lines in all our Mg abundances, but removing these two lines everywhere does not significantly affect our RGB star abundances.

\begin{deluxetable}{lccc}
\tablecolumns{4}
\tabletypesize{\footnotesize}
\tablecaption{\label{tab:corr}Abundance Corrections}
\tablehead{Star & CH Corr. & Na Corr. & Mg Corr.}
\startdata
CarII-6544  & +0.75 & -0.16 & +0.03 \\
CarII-7872  & +0.60 & -0.13 & +0.03 \\
CarII-5664  & +0.74 & -0.23 & +0.05 \\
CarII-0064  & +0.61 & -0.23 & +0.04 \\
CarII-4704  & +0.62 & -0.23 & +0.03 \\
CarII-9296  & +0.49 & -0.26 & +0.04 \\
CarII-2064  & +0.01 & -0.48 & +0.03 \\
CarII-4928  & +0.01 & -0.32 & +0.03 \\
CarII-V3    &\nodata&\nodata&\nodata\\
CarIII-1120 & +0.39 & -0.43 & +0.02 \\
CarIII-8144 & +0.01 & -0.47 & +0.02 \\
\enddata
\end{deluxetable}

Other elements that are known to have significant NLTE corrections include Al, Mn, K, and Fe.
For these elements we do not calculate star-by-star corrections, but instead just estimate the magnitude and direction of a typical correction.
If desired, the effect of these corrections can be approximated by adding the correction to the relevant abundance, as well as adding the total correction in quadrature to the total abundance error; but we do not do so here.

For aluminum, we measured the 3944{\AA} and 3961{\AA} lines, which are heavily affected by NLTE in cool metal-poor stars as well as being in the wings of strong lines, so we only estimate the abundance corrections.
We examined the corrections grid from \citet{Nordlander17}\footnote{\url{https://www.mso.anu.edu.au/~thomasn/NLTE/}} for these lines. Half of our stars are cooler and have lower $\logg$ than the grid range.
The abundance corrections for the 3961{\AA} line tend to be large and positive, from $+0.7$ to $+1.5$ dex.
The corrections for 3944{\AA} are more moderate, from $+0.0$ to $+0.5$ dex. 
The corrections for these lines tend to go in opposite directions, such that averaging corrections for these lines in the warmer stars ($\Teff \gtrsim 4800$\,K) gives corrections in a smaller range from $+0.5$ to $+0.7$ dex.
However, this also tends to make the individual 3944{\AA} and 3961{\AA} abundances more discrepant.
Given these uncertainties, we caution against overinterpretation of our Al abundances or trends.

For manganese, we always use the resonant triplet near 4030{\AA}, as well as redder lines (e.g. 4754\AA, 4783\AA) when detected. \citet{Bergemann19} have recently published grids of Mn corrections, showing overall corrections of about $+0.4$ to $+0.6$ dex, though the corrections are likely larger for cooler and metal-poor stars.
As our Mn abundances just fall within the overall halo trend (which are also not corrected for NLTE), we will not discuss this further.

For potassium, we can measure the 7699{\AA} line in all stars. The 7665{\AA} line was also clear of telluric lines for a few stars, and when measurable is always consistent with the 7699{\AA} line. K has negative NLTE corrections that could be as large as $-0.9$ dex \citep{Ivanova00}, although \citet{Reggiani19} have recently calculated grids of corrections that are more typically $-0.0$ to $-0.4$ dex in our stellar parameter range.

Fe\,I abundances are affected by NLTE effects, with corrections typically $+0.2$ to $+0.3$ dex in our parameter range \citep[e.g.,][]{Bergemann12,Mashonkina16,Ezzeddine17}.
Our temperature correction procedure partially accounts for these effects, though not completely \citep{Frebel13,Ji16d}.
We have decided \emph{not} to apply Fe corrections so as to be able to compare our Fe measurements to literature values, which are essentially all done in LTE.

Finally, we note that Ca can be affected by NLTE as well \citep{Mashonkina16}.
The available grids do not span our whole stellar parameter space\footnote{\url{http://spectrum.inasan.ru/nLTE/}}, but the available corrections are about ${+}0.1$ dex for our stars. We have \emph{not} applied this correction.

\subsection{RRL Abundance analysis}
Stellar parameters for the RRL star CarII-V3 were determined by examining the phase-parameter relations in \citet{For11}.
As our observations are between phases 0.40 to 0.55, stellar parameters are expected to be fairly stable over all exposures.
We adopted initial stellar parameters of $\Teff=6000 \pm 100$\,K, $\logg=1.80 \pm 0.2$\,dex, $\nu_t=3.00\pm0.20\,\kms$ where the error bars are adopted systematic uncertainties based on scatter in the \citet{For11} values.
Then, we measured equivalent widths by fitting Gaussian profiles to the line list from \citet{For10} (rather than our usual line list, which is optimized for red giants).
To slightly improve Fe excitation, ionization, and line strength balance from 28 Fe\,I lines and 10 Fe\,II lines, we adjusted the stellar parameters to $\Teff=6150$\,K, $\logg=1.75$\,dex, $\nu_t=3.15\,\kms$, resulting in $\mbox{[M/H]}=-2.70$.
Total stellar parameter and abundance uncertainties were then determined the same way as the RGB stars.
We do not apply any abundance corrections for this star, as the correction grids are computed for cool giants.
CarII-V3 is one of the most metal-poor RRLs ever studied spectroscopically, with similar [Fe/H] as X Ari and the most Fe-poor RRLs in the LMC \citep{For11,Haschke12,Nemec13}.

\subsection{Abundance Summary}\label{sec:abundsummary}
\begin{figure*}
    \centering
    \includegraphics[width=\linewidth]{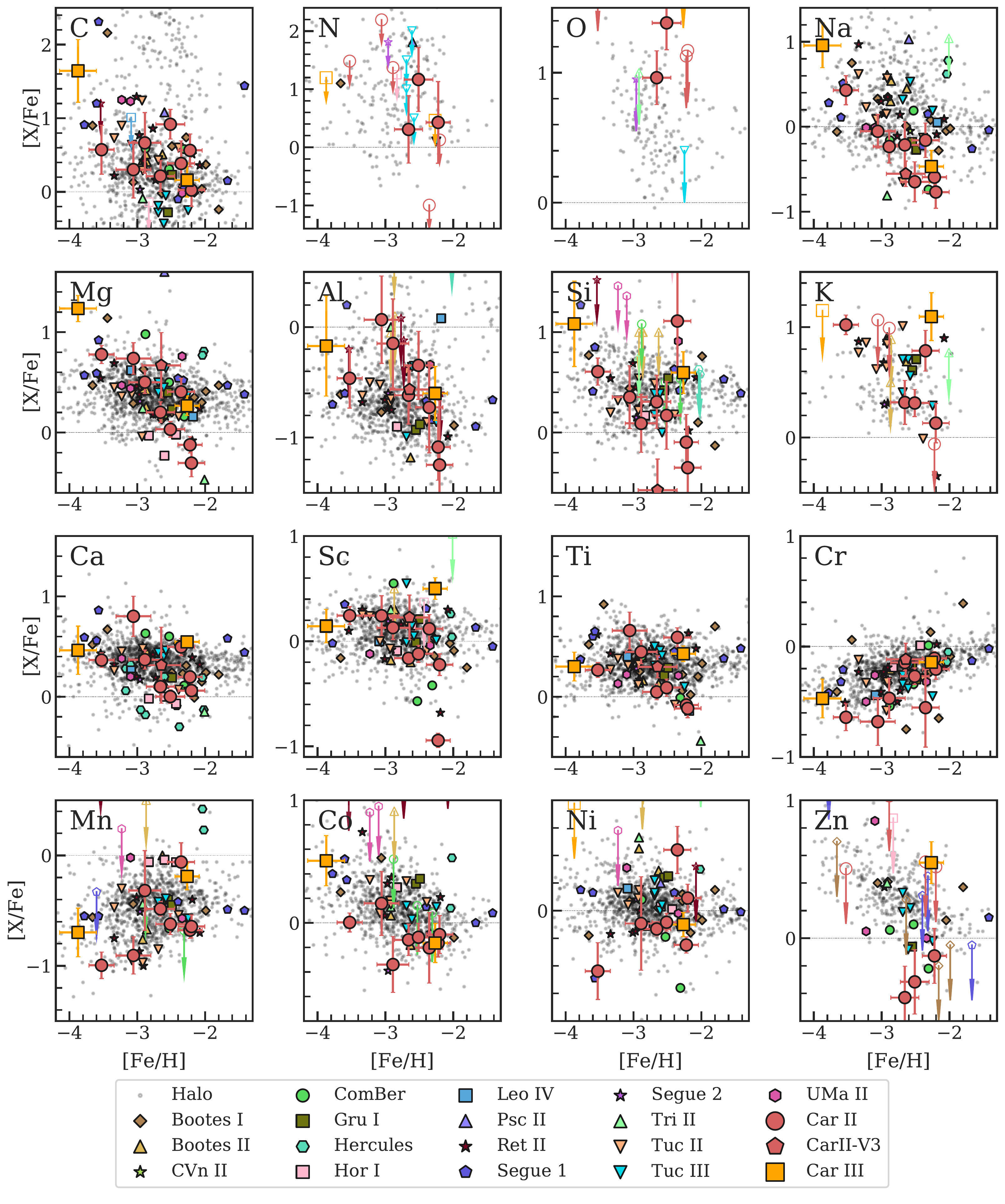}
    \caption{[X/Fe] ratios for most measured elements. Car~II and Car~III are shown as large red circles and large orange squares, respectively, with error bars. 
    The RRL CarII-V3 is shown separately as a red pentagon.
    Other UFDs are shown as small colored points according to the legend.
    Upper limits are indicated by an open point with a downward-pointing arrow.
    The JINAbase halo sample is shown as small grey points in the background.}
    \label{fig:grid}
\end{figure*}

\begin{figure}
    \centering
    \includegraphics[width=0.9\linewidth]{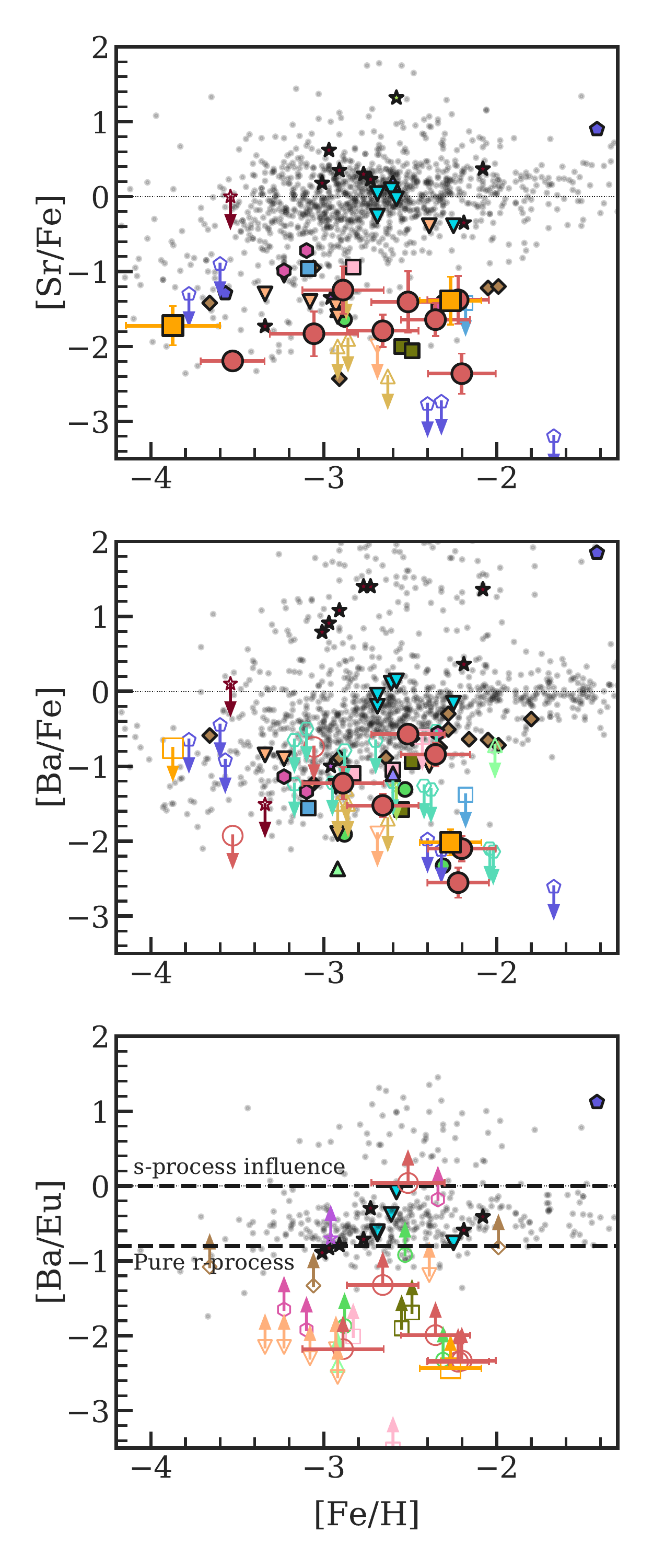}
    \caption{Neutron-capture element abundances in Car~II (large red circles), Car~III (large orange squares), halo stars (grey points), and other UFDs (small colored points, see Figure~\ref{fig:grid} for legend).
    Top and middle panels show [Sr, Ba/Fe]; bottom panel shows [Ba/Eu].
    We draw lines at [Ba/Eu] = $-0.8$ and $0.0$, indicating a pure $r$-process ratio and an $s$-process-influenced ratio \citep{Sneden08}
    Car~II and III match most other UFDs as being deficient in Sr and Ba. Car~II displays significant scatter in [Ba/Fe] at $\mbox{[Fe/H]} \sim -2.5$. One Ba-rich star in Car~II has $\mbox{[Ba/Eu]} > 0$ and thus likely has significant $s$-process enrichment.}
    \label{fig:ncap}
\end{figure}

Our full abundance results are tabulated in Table~\ref{tab:abunds} (Appendix~\ref{app:abunds}) and Figures~\ref{fig:grid} and \ref{fig:ncap}.
We compare the results to halo stars in small grey points \citep{jinabase}, and to other UFD measurements in the literature.
The UFD literature compilation includes Bootes~I \citep{Feltzing09,Norris10a,Gilmore13,Ishigaki14,Frebel16}, Bootes~II \citep{Ji16b}, Canes~Venatici~II \citep{Francois16}, Coma Berenices \citep{Frebel10b}, Grus~I \citep{Ji19a}, Hercules \citep{Koch08, Koch13}, Horologium~I \citep{Nagasawa18}, Leo~IV \citep{Simon10,Francois16}, Pisces~II \citep{Spite18}, Reticulum~II \citep{Ji16c,Roederer16b}, Segue~1 \citep{Frebel14}, Segue~2 \citep{Roederer14a}, Triangulum~II \citep{Ji19a,Kirby17,Venn17}, Tucana~II \citep{Ji16d,Chiti18}, Tucana~III \citep{Hansen17,Marshall18}, and Ursa~Major~II \citep{Frebel10b}.
We reiterate that throughout this paper, the error bars for Car~II and III include full propagation of the line-by-line statistical and stellar parameter uncertainties.

The RRL star CarII-V3 generally has consistent abundances with the RGB stars, although there are fewer lines and only moderate S/N so the abundance uncertainties for this star are fairly large.
The main outlier is the Si abundance, which is unusually low but has large uncertainty as it is measured only from the 3905{\AA} line.
Given the abundance similarities to other stars in Car~II, we will treat this star's abundances on the same footing as RGB stars when lines are detected.

\emph{C, N, O.} Carbon abundances are derived from synthesizing the CH bands at ${\sim}4300-4325$\AA. CO molecular equilibrium affects CH abundances, and we always assume the MOOG default of [O/Fe]$=0$ even when O is measured independently.
Literature measurements suggest [O/Fe] is typically $>0.5$ \citep[e.g.,][]{Brown14}. If we used [O/Fe]$=+1.0$ instead, [C/Fe] would typically increase by ${+}0.08$ dex with star-to-star scatter of 0.08 dex, but we keep the MOOG default for consistency with previously analyzed literature stars.
Nitrogen is derived from fitting CN bands at ${\sim}3850$\AA\ after fixing the CH abundance.

In two relatively cool and metal-rich stars, we detect the two forbidden oxygen lines at ${\sim}6300$\AA. These can only be measured when the O abundance is very high, so are probably a biased sample of measurements. 
The stronger 6300\AA\ line was deblended from telluric absorption, and the weaker 6363\AA\ line can be affected by a wide calcium ionization feature \citep[e.g.,][]{Barbuy15}. However in both cases, the two different lines give very close abundances.
We include oxygen upper limits for all stars (including the two detections) in the machine-readable version of Table~\ref{tab:lines} from the 6300\AA\ line.

\emph{$\alpha$-elements: Mg, Si, Ca.} The $\alpha$-element abundances are determined from equivalent widths in all stars. Magnesium is determined from 5-7 lines including the Mg b lines in all stars (except CarII-V3, where only the Mg b lines can be measured). The Mg b lines are quite strong and saturated but give similar abundances as the weaker lines for all stars.
Si is measured from both the 3905{\AA} and 4102{\AA} lines, but these are both rather poor-quality lines. The 3905{\AA} line is fairly saturated, and the 4102{\AA} line is in a Balmer wing.
Ca is usually measured from 10-20 lines with three exceptions: the warmer and more Fe-poor stars CarII-4928 and CarIII-1120 have only 2 and 1 Ca lines, respectively; and only the strong 4226{\AA} line is detected in the RRL CarII-V3. We do not use the 4226{\AA} line in any of the RGB stars due to large and uncertain NLTE corrections \citep[e.g.,][]{Sitnova19}.

\emph{Odd-Z elements: Na, Al, K, Sc.}
We use equivalent widths to measure sodium abundances from the two Na D lines, which have been corrected for NLTE effects.
We synthesize the 3944{\AA} and 3961{\AA} Al lines, which are both very strong and subject to NLTE effects so our Al abundances are very uncertain.
K abundances are mostly from the 7699{\AA} line, although occasionally the 7665{\AA} is not blended with tellurics.
Sc abundances are mostly measured with spectral synthesis from five lines at $4246 < \lambda < 4415${\AA}, though the redder line abundances (e.g. 5031\AA, 5526\AA) agree.

\begin{figure}
    \centering
    \includegraphics[width=\linewidth]{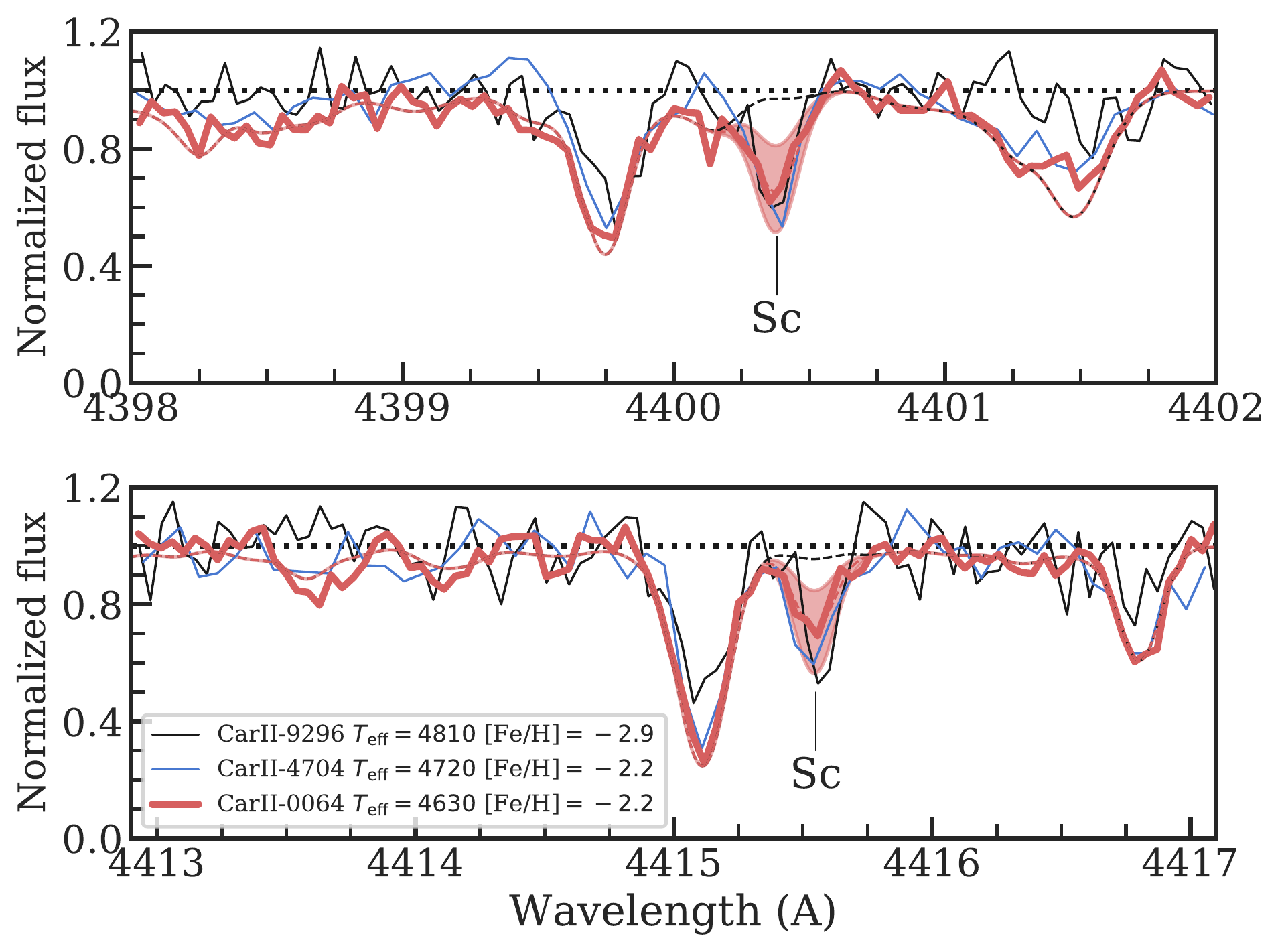}
    \caption{Spectrum of the low-Sc outlier CarII-0064 around two Sc lines, compared to two other Car~II stars with similar temperature but lower Sc abundance.
    The Sc line is deficient in CarII-0064 compared to these other stars despite this star being somewhat cooler.
    Note that there is C-H absorption in CarII-0064 near the 4400\AA\ line. We also show the synthetic spectrum fit to the Sc line for CarII-0064 as a thin dashed red line, a $\pm 0.5$ dex difference to the synthetic fit as a shaded region, and a synthesis with no Sc as a dashed black line.
    }
    \label{fig:scandium}
\end{figure}

CarII-0064 is a significant low Sc outlier in Car~II with [Sc/Fe] $\approx -1$ (Figure~\ref{fig:grid}). We plot two Sc line spectrum in Figure~\ref{fig:scandium}, along with its synthetic fit and two other stars that have higher Sc abundances. The Sc abundance is clearly lower in CarII-0064, though visually not as much as would be expected from Figure~\ref{fig:grid}.
This is because each individual line difference is significant at $\lesssim 2\sigma$, but they are all consistent and the combination of $5-6$ Sc lines reduces the uncertainty. Also note the [Sc/Fe] abundance error is smaller, due to correlated uncertainties in stellar parameters.
Such low Sc abundances have previously been seen in ``iron-rich'' stars (those with overall low [X/Fe] ratios, e.g., \citealt{Cohen10,Cohen13,Yong13a}).
However, this cannot explain CarII-0064 because it is an outlier from the overall Car~II trend only in [Sc/Fe].
Similarly Sc-deficient stars have been found in the bulge where it has been argued that this signature may indicate unusually old stars \citep{Casey15}, but we see no sign of this in the more Fe-poor stars in Car~II.
It is unclear to us how to interpret this star's extreme Sc abundance.

\emph{Fe-peak elements: Ti, Cr, Mn, Co, Ni, Zn.}
We use equivalent widths to measure abundances for both ionization states of titanium, but we adopt the Ti\,II abundances everywhere as our default; it is measured in all our stars, has more and stronger lines, and is less susceptible to NLTE effects.

The Fe-peak elements closely follow the halo trends within their abundance uncertainties.
There are minor deviations that are all significant at $<2\sigma$, so we do not concern ourselves with these further, other than to comment that Zn could be moderately enhanced in Car~III and moderately deficient in Car~II.

\emph{Neutron-capture elements: Sr, Ba.} These elements have low abundances or upper limits, similar to most other UFDs.
The nucleosynthetic origin of these very low Sr and Ba abundances remains unknown (it is in general not even clear if they are from the $r$- or $s-$processes, see \citet{Ji19a} for an extensive discussion), but it appears to be unique to UFDs and occasional halo stars that are presumably stripped from UFDs.
Given the low abundance of neutron-capture elements, no other neutron-capture elements could be detected, so we place [Eu/Fe] upper limits and show [Ba/Eu] in Figure~\ref{fig:ncap}.

There are two stars in Car~II with relatively high $\mbox{[Ba/Fe]} \gtrsim -1$ compared to the other Car~II stars.
One of these relatively Ba-rich stars, CarII-7872, also has a low Eu upper limit that results in $\mbox{[Ba/Eu]} \gtrsim 0$, suggesting its Ba is predominantly from the $s$-process \citep[e.g.,][]{Sneden08}.
We discuss this large barium scatter in Section~\ref{sec:bascat}.

\section{Formation history of Carina~II and III}\label{sec:formation}
\subsection{Carina II and III are Dwarf Galaxies}
Low luminosity stellar systems are classified as either dwarf galaxies or star clusters. Dwarf galaxies are generally more spatially extended than clusters, with velocity dispersions implying significant dark matter content and nonzero metallicity (or more specifically, iron-peak abundance) dispersions \citep{Willman12}.
Faint dwarf galaxies also tend to display very low abundances of neutron-capture elements \citep[e.g.,][]{Ji19a}, while globular clusters have light element anticorrelations associated with hot bottom burning \citep[e.g.,][]{Bastian18}.

Both Carina~II and III are clearly dwarf galaxies and not globular clusters.
Their half-light radii and luminosities place them within the dwarf galaxy morphological locus \citep{Torrealba18}.
Carina~II displays both a significant velocity and metallicity dispersion from medium-resolution data \citep{Li18Carina}.
Our two Carina~III stars have [Fe/H] values that differ by almost 2 dex, definitively establishing a significant metallicity dispersion. We have also now resolved the velocity dispersion \citep{LiPrep}.
The neutron-capture elements Sr and Ba are low in both systems, like nearly every other UFD (Figure~\ref{fig:ncap}).

\begin{figure}
    \centering
    \includegraphics[width=0.7\linewidth]{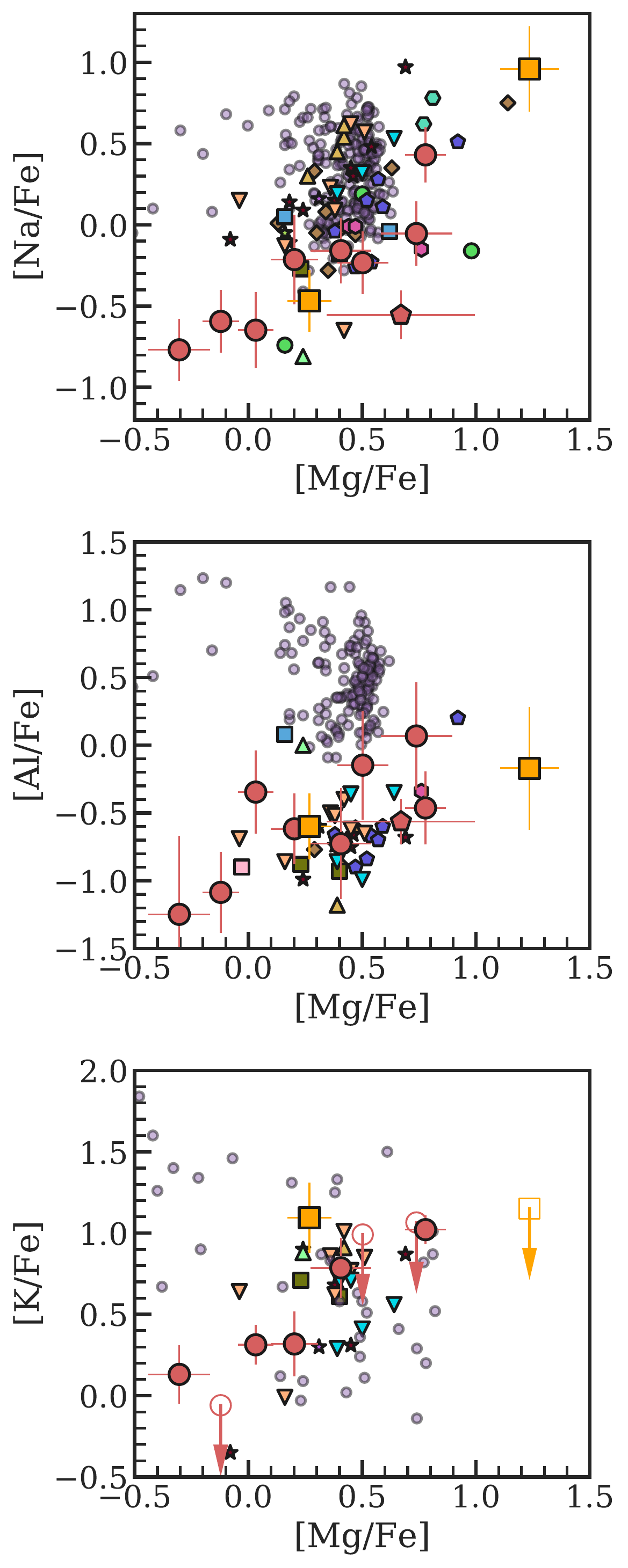}
    \caption{Mg-Na, Mg-Al, and Mg-K abundance patterns. Car~II is red circles/pentagon, Car~III is orange squares, globular cluster stars are small purple circles, and other UFDs are small colored points (same as Figure~\ref{fig:grid}).
    In globular clusters, Mg-Na and Mg-Al are \emph{anti}-correlated, while in both Car~II and Car~III these elements are clearly correlated.
    Mg-K are anticorrelated in the globular cluster NGC 2419, and there is no evidence for such in Car~II.
    The light element correlations confirm that Car~II and III are dwarf galaxies and not globular clusters.
    }
    \label{fig:gcabund}
\end{figure}

These criteria alone already show that Car~II and III are galaxies, but as a final confirmation we show there are no light element anticorrelations.
Figure~\ref{fig:gcabund} shows these relations for our stars.
In the top two panels, we show Na-Mg and Al-Mg for our UFD stars (symbols as in Figure~\ref{fig:grid}) and globular cluster stars as purple circles \citep[from references][]{Carretta07,Carretta09,Gratton06,Cohen12}.
Most globular clusters do not show significant dispersion in [Mg/Fe], but those that do always display an \emph{anti}-correlation in Na-Mg and Al-Mg.
In contrast, there is very clearly a positive correlation for these elements in both Car~II and III.
Note that Na and Mg have NLTE corrections applied, while the Al corrections should on average provide an offset and are unlikely to turn a strong positive Mg-Al correlation into an anticorrelation.
The bottom panel of Figure~\ref{fig:gcabund} shows the Mg-K anticorrelation found in NGC 2419 \citep{Mucciarelli12}, which is not present in Car~II. However, our two stars in Car~III (including one K upper limit) do not rule out an Mg-K anticorrelation in this system.

\subsection{Car~II and III are consistent with being accreted along with the LMC/SMC}
\citet{Li18Carina} showed that the positions and radial velocities of both Car~II and Car~III were consistent with having accreted with the LMC, according to the \citet{Jethwa16} model.
\citet{Kallivayalil18} then added proper motion data from Gaia DR2 \citep{GaiaDR2,GaiaSatellite}, finding that Car~II and Car~III are also likely LMC satellites based on LMC analogues in the Aquarius simulations (\citealt{Springel08}, also see \citealt{Sales17,Simon18,Erkal19}).
Kinematically, it thus appears likely that both Car~II and Car~III entered the Milky Way with the LMC/SMC system, although Car~II is towards the edge of the likely region due to its high velocity.
\citet{Kallivayalil18} also associate Hyi~I and Hor~I with the LMC.

Thus, Car~II and III, along with Hor~I \citep{Nagasawa18}, can be studied in contrast to other UFDs to see if abundance ratios have any environmental dependence.
\citet{Nagasawa18} point out that the three stars in Hor~I have unusually low Mg and Ca, with one possible explanation being that LMC satellites might have typically different enrichment histories compared to Milky Way UFDs.
Figure~\ref{fig:grid} does not suggest that Car~II or Car~III obviously deviate from the typical abundance scatter of other UFDs, including for Mg and Ca. The unusually low Mg and Ca in Hor~I thus likely has some other origin.

\subsection{$\alpha$-element evolution: time delay scenario or initial mass function variations?}\label{sec:alphaevol}

\subsubsection{$\alpha$-element abundance ratios in Car II and Car III}
The $\alpha$-elements (O, Mg, Si, Ca)
are primarily produced in core-collapse supernovae (CCSNe) and thus tend to be enhanced at low [Fe/H].
After a delay of $100-1000$ Myr \citep{Maoz14}, Type~Ia supernovae (SNe1a) begin to add Fe peak elements, causing a ``knee'' in [$\alpha$/Fe] vs [Fe/H] \citep{Tinsley79}.
In this time delay scenario, the location of the knee can be interpreted as an overall star formation timescale for a galaxy \citep[e.g.,][]{Tolstoy09,Kirby11}.
Figure~\ref{fig:grid} shows clear downward trends in [Mg/Fe] and [Ca/Fe] vs. [Fe/H] for both Car~II and III, with a possible knee at $\mbox{[Fe/H]} \sim -2.8$ for Car~II that would indicate very slow chemical evolution in this low mass galaxy.

\begin{figure}
    \centering
    \includegraphics[width=\linewidth]{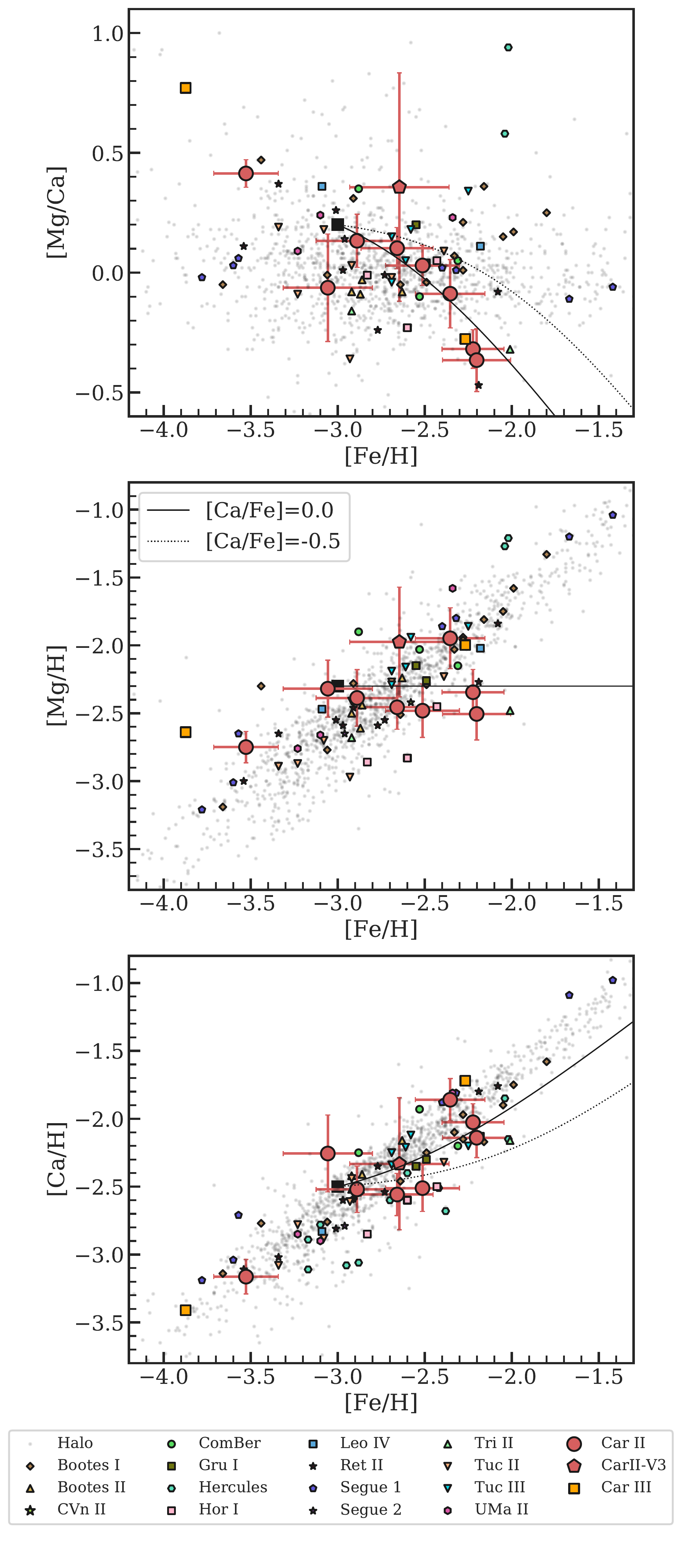}
    \caption{Top panel: [Mg/Ca] vs [Fe/H] for UFDs (colored points) and halo stars (small grey points). We focus here particularly on Car~II (large red points with error bars).
    Middle and bottom panels: [Mg/H] and [Ca/H] as a function of metallicity.
    On all panels, the solid and dotted black lines show tracks of SN1a-only enrichment for two SN1a Ca yields, starting at the black square.
    }
    \label{fig:mgca}
\end{figure}

However, there is a striking difference in the size of the trend for [Mg/Fe] and [Ca/Fe]: [Mg/Fe] declines by over 1 dex, while [Ca/Fe] declines by only about 0.4 dex.
We will focus primarily on Car~II, because Car~III has only two stars and the more Fe-poor star has only one Ca line.
To clarify the Mg and Ca difference, in the top panel of Figure~\ref{fig:mgca} we plot [Mg/Ca] vs [Fe/H], where [Mg/Ca] declines from about $+0.4$ to $-0.4$ as [Fe/H] increases from $-3.5$ to $-2.2$.
These extreme [Mg/Ca] ratios are often interpreted as variations in the high mass end of the initial mass function.
Stars with high [Mg/Ca] ratios are typically associated with enrichment by very massive stars with $M > 20-30 M_\odot$ \citep[e.g.,][also see Section~\ref{sec:hw10fit}]{Norris00,Cohen07,Koch08}.
Stars with $\mbox{[Mg/Ca]} < 0$ form out of gas enriched by lower mass CCSN progenitors with $M \lesssim 15 M_\odot$ \citep[e.g.,][]{Tolstoy03,McWilliam13}.
The variable [Mg/Ca] ratios in Car~II may thus indicate that the $\alpha$-elements in this galaxy is tracing changes in the high-mass end of the initial mass function (IMF).
Indeed, the low-mass end of the IMF in UFDs has previously been shown to vary between different UFDs \citep{Geha13,Gennaro18}, which tantalizingly hints that the high-mass end of the IMF might vary as well (although the low-mass IMF varies from galaxy to galaxy, while here we consider time variations within a single galaxy, so the mechanisms may not be related).

In the bottom panels of Figure~\ref{fig:mgca}, we plot [Mg/H] and [Ca/H] vs [Fe/H], which shows that there may actually be two phases of [Mg/Ca] evolution: from $\mbox{[Fe/H]} = -3.6$ to $-3.0$ this is primarily driven by a smaller increase in [Mg/H] than [Ca/H]; while from $\mbox{[Fe/H]} = -3.0$ to $-2.2$, [Mg/H] stays mostly flat while [Ca/H] increases.
The first phase unambiguously shows that Car~II has been enriched by at least two different masses of CCSNe: the most Fe-poor star in Car~II has high [Mg/Ca] ratios suggesting enrichment by high mass stars, but it has lower [Mg/H] than the higher metallicity stars. Since SNe1a produce negligible Mg, this means that CCSNe with $\mbox{[Mg/Ca]} \sim 0$ must have enriched Car~II after the formation of the most Fe-poor star.
This could potentially be evidence of a transition from very massive Pop~III stars to regular mass Pop~II CCSNe.

The second phase of evolution could be attributed to either IMF variation or SN1a enrichment.
To illustrate this, we show an extremely simple chemical evolution track in Figure~\ref{fig:mgca}. First, we set an initial [Mg/H], [Ca/H], and [Fe/H] that matches the [Mg/Ca] ratio at $\mbox{[Fe/H]}=-3$ (black square).
Then, we assume a fixed [Ca/Fe] yield and negligible Mg yield for SNe1a \citep{Kirby19}, and compute the evolution of Mg, Ca, and Fe assuming no more CCSNe and no gas accretion/expulsion.
\citet{Kirby19} have recently made an empirical measurement of the SN1a [Ca/Fe] yield in larger dSph galaxies, finding values that range between $-0.5 < \mbox{[Ca/Fe]} < 0.0$.
We thus apply our simple model with SN1a yields of [Ca/Fe] $= 0.0$ and $-0.5$, which are shown as black solid and dotted lines respectively in Figure~\ref{fig:mgca} and reasonably match the observed Mg and Ca ratios.
This would be quite an extreme situation: if most of the metal enrichment in Car~II is due to SNe1a and not CCSNe, but stars still formed to sample the SN1a yields, that implies an extremely top-light IMF where no massive stars formed.
However, this is definitely not a unique model, and specifically the flat [Mg/H] trend does not rule out contributions from additional CCSNe because gas accretion can increase the hydrogen reservoir \citep[e.g.,][]{Ji16b}.
Detailed chemical evolution modeling of more elements might help clarify the picture but is beyond the scope of this paper.
Furthermore, stochastic sampling of individual SN explosions may dominate the observed trends \citep[e.g.,][]{Koch08,Koch13,Revaz16,Applebaum18}, especially given that Car~II produced only ${\sim}100$ CCSNe in total (assuming a Salpeter initial mass function and present-day mass-to-light ratio of 2.2, \citealt{Ji16b}). Car~III is even more susceptible to stochastic enrichment, having been enriched by only ${\sim}15$ supernovae. We thus caution against over-interpreting the available data.

\subsubsection{[Mg/Ca] abundances across the UFD population}

Some more insight can be derived by comparing the [Mg/Ca] vs [Fe/H] trends of Car~II to the trends in other UFDs.
It turns out that few other UFDs have similarly negative [Mg/Ca] vs [Fe/H] slopes. 
To quantify this result, we fit lines to the [Mg/Ca] vs [Fe/H] evolution of every UFD individually, and consider the slope \emph{angle} (i.e., $0^\circ$ corresponds to a flat line, and negative slope angles indicate declining [Mg/Ca] as [Fe/H] increases).
We then calculate the slopes and slope uncertainties by assuming that data points are drawn from a thin line with multivariate Gaussian uncertainties \citep[see section 7 of ][]{Hogg10}.
We take a uniform prior in slope \emph{angle} (as opposed to slope) for $\theta \in [-90^\circ, +90^\circ)$ and a flat prior for the intercept, then use \texttt{emcee} to sample the posterior \citep{emcee}.
We take the posterior median as the point estimate and the 16th-84th percentile range as the 68\% credible interval.
We remove the four UFDs that have unconstrained posteriors (since their stars have essentially the same [Fe/H]).
Note that the literature UFD stars have inhomogenously determined uncertainties, so we instead assume independent error bars of 0.2 dex for both [Fe/H] and [Mg/Ca]; but use our actual abundance uncertainties for Car~II and Car~III.

\begin{figure}
    \centering
    \includegraphics[width=\linewidth]{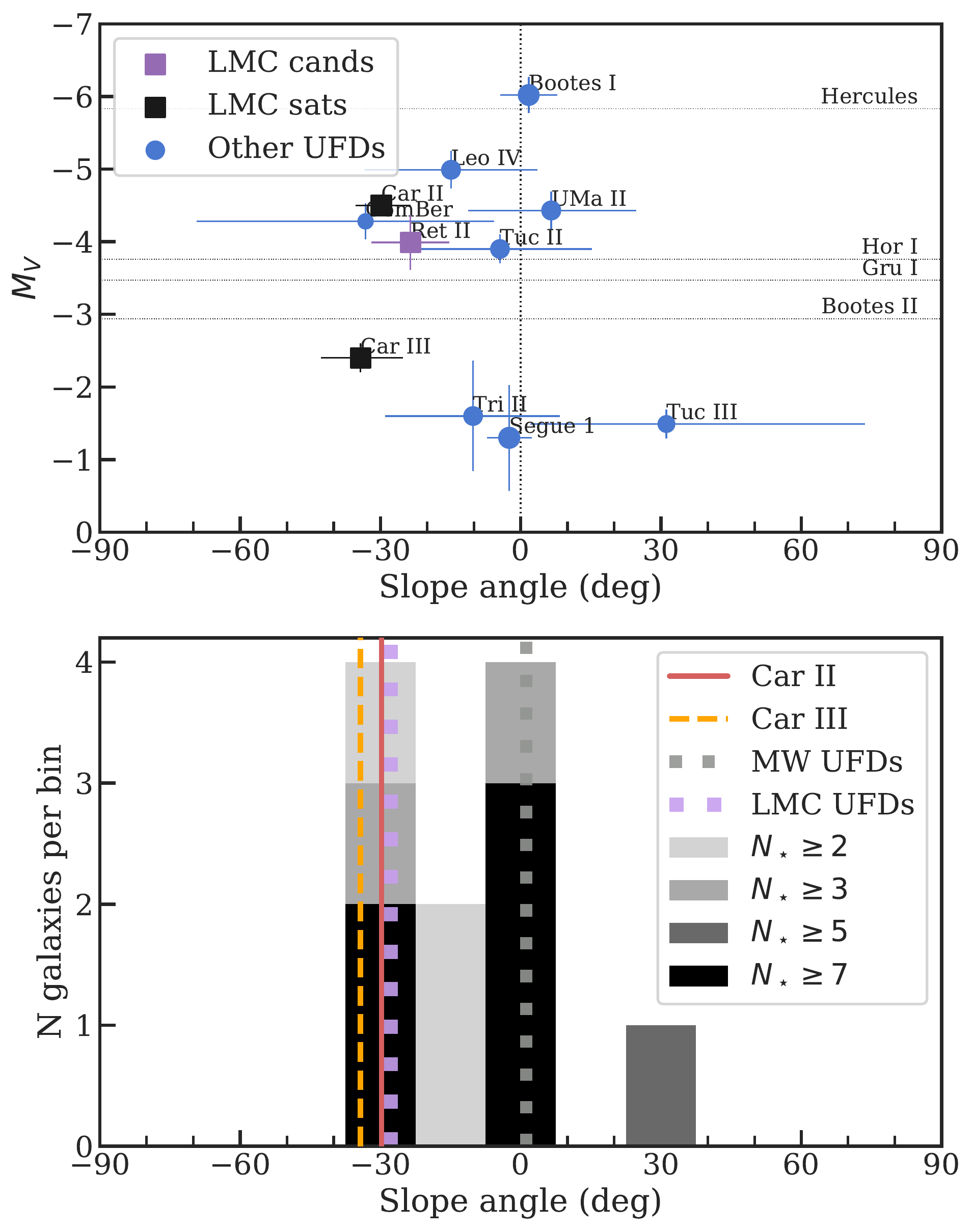}
    \caption{Top: [Mg/Ca] vs [Fe/H] slope angles for all UFDs vs luminosity $M_V$.
    Slope error bars indicate 68\% posterior region, and measurements with smaller uncertainties have correspondingly larger symbols.
    The luminosities of galaxies with unconstrained slope posteriors are shown as horizontal lines.
    Square symbols indicate UFDs that are LMC satellites (Car~II, Car~III, Hor~I) or satellite candidates (Ret~II).
    Round symbols indicate UFDs likely associated with the Milky Way.
    Bottom: histogram of [Mg/Ca] slope angles for all UFDs. UFDs with more stars (i.e., more confident slope measurements) are shown as darker shades of grey. The relatively extreme slope angles for Car~II and III are marked as vertical solid and dashed lines, respectively. The total [Mg/Ca] slopes for all MW vs LMC UFD stars are marked in dotted grey and purple lines, respectively.
    }
    \label{fig:mgcaenv}
\end{figure}

The [Mg/Ca] vs [Fe/H] slopes for all UFDs where ${\geq}2$ stars have detailed abundance measurements are shown in Figure~\ref{fig:mgcaenv}.
The top panel of Figure~\ref{fig:mgcaenv} shows the UFD [Mg/Ca] slopes vs luminosity (luminosities from the \citealt{Simon19} compilation, including data from \citealt{Bechtol15,Munoz18,Torrealba18,MutluPakdil18}).
There is not an obvious relation between slope angle and luminosity.
The bottom panel shows a histogram of the slope angle point estimates from the top panel.
Many UFDs have too few stars to place a useful slope constraint, so we shade each UFD in the histogram by the number of stars used to calculate the slope, with darker colors indicating more stars. The UFDs with the most confident measurements (i.e., ${\geq} 7$ stars with detailed abundances) are Car~II (this work), Ret~II \citep{Ji16c}, Bootes~I \citep{Frebel16}, Segue~1 \citep{Frebel14}, and Tuc~II \citep{Chiti18}.
We also highlight the slope of Car~II and Car~III as a vertical solid red line and vertical orange dashed line, respectively.
Of the other UFDs, only Ret~II exhibits a declining [Mg/Ca] slope that deviates from zero by $\gtrsim1\sigma$.

\subsubsection{Effect of environment on [Mg/Ca] abundances}
The results above raise an interesting question about the role of environment in determining abundance trends:
Car~II and III are LMC satellites, and Ret~II is also a candidate LMC satellite 
\citep{Kallivayalil18,Erkal19}\footnote{Hor~I \citep{Nagasawa18} also is an LMC satellite, but all three currently observed stars have $\mbox{[Fe/H]} \sim -2.6$ within uncertainties and thus no useful constraint on its [Mg/Ca] vs [Fe/H] trend. The three Hor~I stars all have $\mbox{[Mg/Fe]} \approx \mbox{[Ca/Fe]} \approx 0$.}.
In the bottom panel of Figure~\ref{fig:mgcaenv}, we show the [Mg/Ca] vs [Fe/H] slope angles from grouping all LMC UFD stars and all MW UFD stars.
It is very obvious that the LMC satellite UFD stars have a significant negative slope, while the MW satellite UFD stars have a flat slope; though we note that the LMC trend is mostly driven by Car~II and should await additional abundances in LMC satellite UFDs to clarify this suggestion.

However, we speculate briefly on how the large scale environment could possibly affect chemical evolution in UFDs.
At first glance, UFDs should \emph{not} display significant environment dependence.
UFDs form most of their stars by $z \sim 6$ \citep{Brown14}, and in simulations the closest more massive galaxy at $z>6$ is typically 400 physical kpc away (\citealt{Wetzel15}).
Even generously sized galactic superbubbles reach only tens of kpc \citep{Griffen18}, so external enrichment or directly affecting UFD gas with ram pressure stripping is unlikely \citep{Wetzel15}.
However, radiation (both ionizing and Lyman-Werner) can span these distances, though there are limited ways we can imagine this would affect stellar populations.
At the metal-rich end, one possibility is the integrated galactic IMF theory \citep[IGIMF, e.g.,][]{Weidner13,McWilliam13}, which suggests that as galaxies become gas-poor they cannot form the most massive stars. If LMC UFDs formed later and thus reionized earlier in their evolution, they would form more of their stars in this phase.
At the metal-poor end, delaying Pop~III star formation with Lyman-Werner feedback may increase susceptibility of UFD progenitors to external enrichment \citep[e.g.,][]{Magg18}.
Also, metal-free gas with relatively high ionization fractions can form HD molecules during collapse, which may (or may not) affect the Pop~III initial mass function \citep{Glover13}.
A final note is that the distance scales from \citet{Wetzel15} assume that UFDs reside in dark matter halos of $M_{\rm peak} \sim 10^9 M_\odot$ \citep{Wetzel15}.
If instead UFDs reside in smaller dark matter halos of $M_{\rm peak} \sim 10^{7-8} M_\odot$ (e.g., \citealt{Jeon14, Ji15, Jethwa18, Graus19}), then separation distances would become smaller and environmental effects could be more important.

\subsection{Inhomogenous metal mixing of AGB winds in Car~II}\label{sec:bascat}
There is real scatter in [Ba/Fe] at $\mbox{[Fe/H]} \sim -2.5$ in Car~II, with some stars having relatively high Ba abundances and others having low Ba abundances (Figure~\ref{fig:ncap}). The extent of the scatter in Ba is ${\sim}1$ dex, much larger than the scatter in any other abundance ratio.
A plausible explanation for the Ba scatter is inhomogeneous mixing of AGB wind ejecta into the galaxy's ISM. Unlike supernova ejecta, which mix rapidly upon entering the hot phase of the ISM, AGB winds mix into relatively cool ISM phases and can thus stay quite inhomogeneous \citep{Emerick18, Emerick19}.
Since Ba is produced by the $s$-process and released in AGB winds, this mechanism could explain the large Ba scatter.
This scenario is supported by the fact that one of the high-Ba stars (CarII-7872) has $\mbox{[Ba/Eu]} \gtrsim 0$ (Figure~\ref{fig:ncap}), suggesting its Ba is predominantly from the $s$-process.
Since most barium comes from AGB stars with initial mass $M \leq 4 M_\odot$ and lifetimes $\geq 10^8$ years, the presence of AGB enrichment requires that Car~II formed stars for at least ${\sim}100$ Myr \citep{Lugaro12,Karakas16}.
Note that the nucleosynthetic origin of the low Sr and Ba floor in UFDs remains unknown \citep[see][for more discussion]{Ji19a}.
One might also expect a correlation between Ba and other AGB elements like C. We find a moderate but not statistically significant correlation between stars that have both Ba and C detected in Car~II (correlation of $0.48$ with a $p$-value of $0.34$ from \texttt{scipy.stats.pearsonr}).

\section{Population~III Star Signatures}\label{sec:pop3}
\subsection{Carbon-enhanced fraction in UFDs}
Carbon-Enhanced Metal-Poor (CEMP) stars are stars with high [C/Fe] ratios \citep{Beers05}.
Below $\mbox{[Fe/H]} \sim -3$, about half the stars in the Milky Way halo are CEMP stars (i.e., $\mbox{[C/Fe]} \gtrsim +0.7$, \citealt{Aoki07}).
It is generally thought that a specific subclass (CEMP-no stars; \citealt{Beers05})\footnote{The ``\emph{no}'' is short for ``no strong enhancement of neutron-capture elements".} of the CEMP stars traces unique nucleosynthesis in Pop~III stars \citep[e.g.,][]{Norris13,Frebel15,Placco16}.
If so, the observed CEMP fraction provides a window to the distribution of some Pop~III star properties, such as initial mass, explosion energy, or stellar rotation \citep[e.g.,][]{Cooke14,Ji15}.

\begin{figure}
    \centering
    \includegraphics[width=\linewidth]{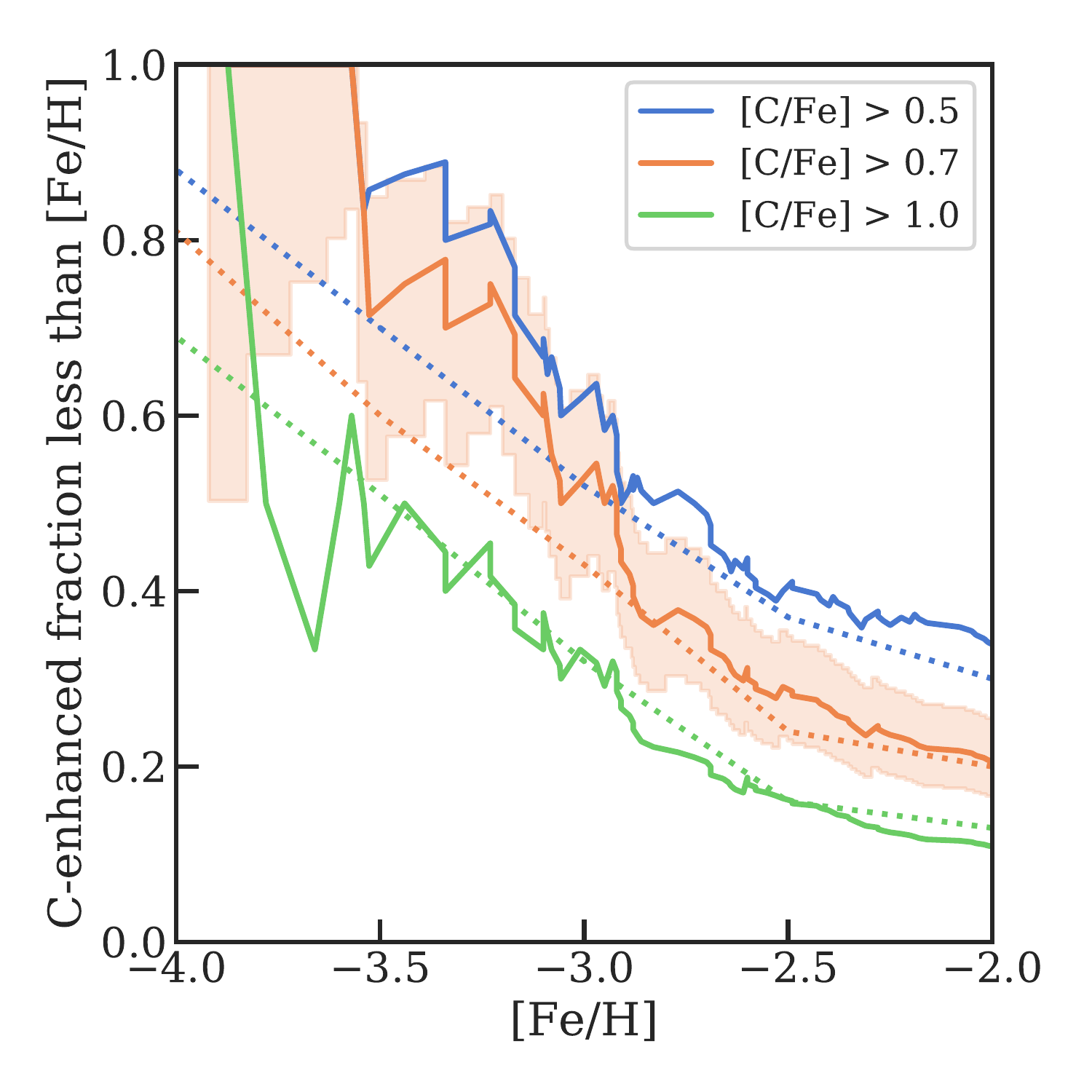}
    \caption{Cumulative carbon-enhanced metal-poor (CEMP) fraction.
    The solid colored lines show the CEMP fraction for UFD stars at different C-enhanced cutoffs, with the shaded region indicating the 68\% Wilson confidence interval for a binomial distribution around the ${\rm [C/Fe]}>0.7$ fraction.
    The dotted colored lines show the halo CEMP fraction from \citet{Placco14}.
    Both the UFD data and the reference sample have included carbon evolutionary corrections.
    The UFD CEMP fraction is consistent with the halo.
    }
    \label{fig:cfrac}
\end{figure}

In Figure~\ref{fig:cfrac}, we show the fraction of carbon-enhanced stars below a given [Fe/H] in our ${\gtrsim}80$ star UFD literature sample and the halo star compilation by \citet{Placco14}.
Both samples have included the \citet{Placco14} evolutionary carbon corrections.
For the UFD sample, we show 68\% Wilson confidence intervals on the CEMP fraction.
Figure~\ref{fig:cfrac} shows that the carbon-enhanced fraction in UFDs is essentially identical to halo stars at all levels of carbon enhancement.
For comparison, the CEMP fraction in larger dwarf galaxies like Sculptor has been studied in some detail \citep[e.g.,][]{Skuladottir15,Salvadori15,Chiti18b}, but it is still debated whether the CEMP fraction in those galaxies is consistent with the halo.

If we are after pure Pop~III signatures, it also makes sense to look at entire UFDs as either C-rich or C-normal \citep{Ji15}.
Seven UFDs have stars with $\mbox{[Fe/H]} < -3$. The most metal-poor stars in five of these UFDs are C-rich (Car~III, Segue~1, Boo~I, Tuc~II, UMa~II), while the other two are C-normal (Ret~II, Car~II).
This suggests that the fraction of Pop~III stars producing carbon-enhanced abundances is $0.71_{-0.19}^{+0.13}$, following the simple models in \citet{Ji15}.
A more stringent cut of $\mbox{[Fe/H]} < -3.5$ results in three C-enhanced galaxies out of five, or a carbon-enhanced rate of $0.60^{+0.34}_{-0.39}$.
More to the point, the existence of carbon-normal stars with $\mbox{[Fe/H]} \lesssim -3.5$ in Ret~II and Car~II is evidence against the hypothesis that 100\% of Pop~III stars produce carbon-enhanced signatures, as is often assumed in theoretical models and simulations \citep[e.g.,][]{Salvadori15, Jeon17}. 

\subsection{Full fits to individual UFD stars}\label{sec:hw10fit}

\begin{figure}
    \centering
    \includegraphics[width=.79\linewidth]{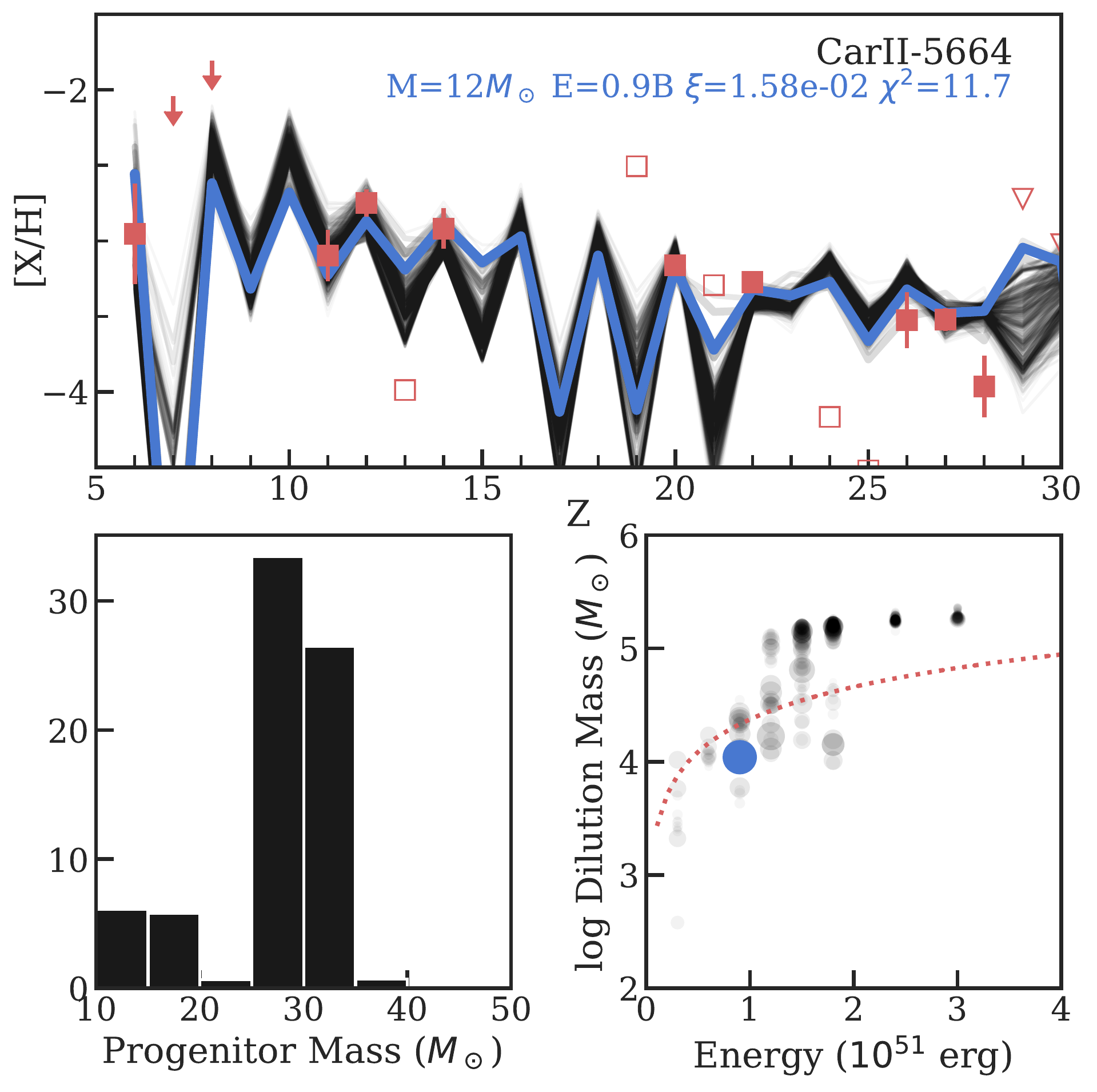}\\[5mm]
    \includegraphics[width=.79\linewidth]{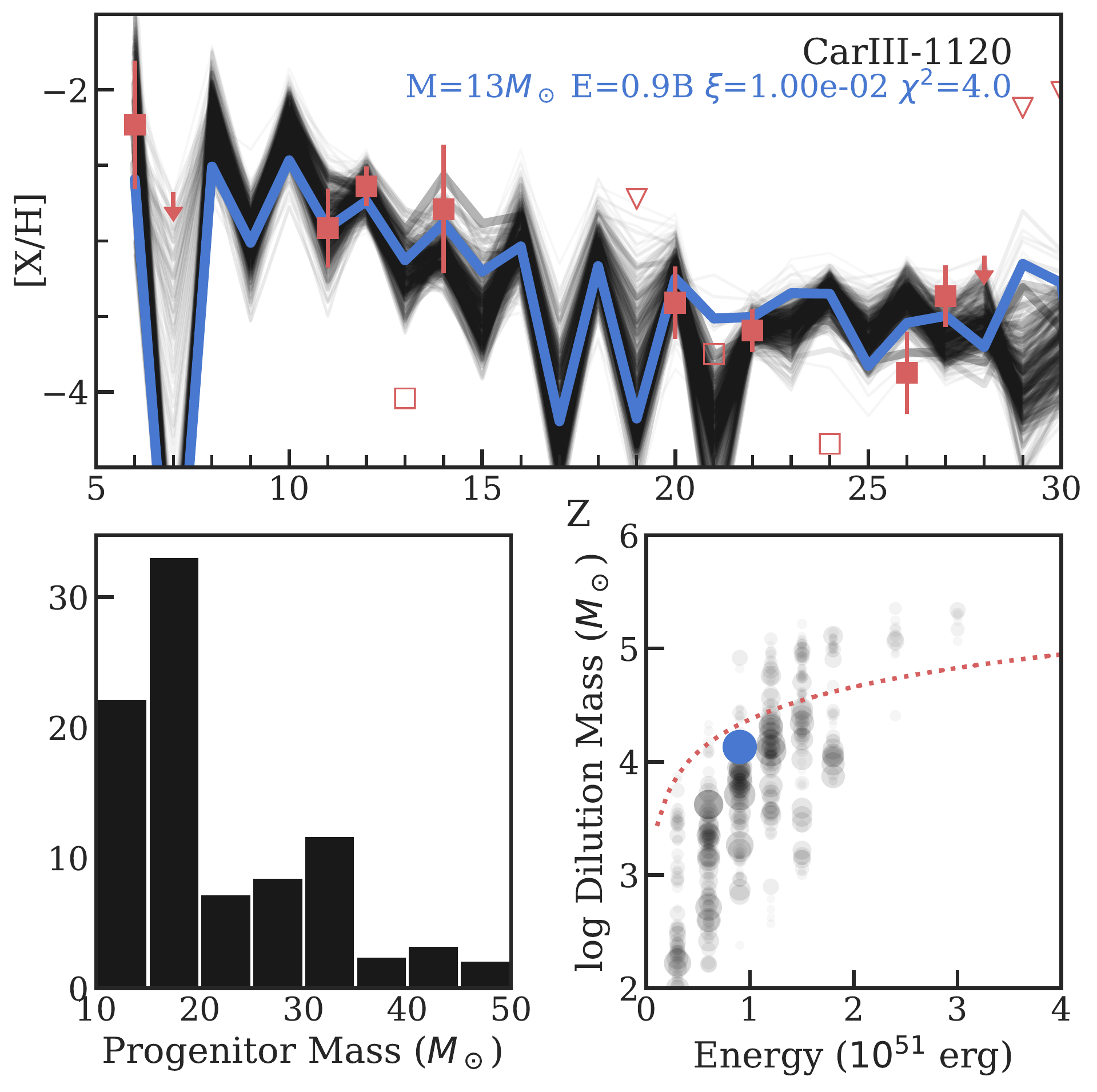}
    \caption{Pop~III SN yield fits to abundances of the three stars with $\mbox{[Fe/H]} \lesssim -3.5$. 
    \emph{Top panels:} the measured abundances (red), the single best-fit model (blue) and all models within 2$\sigma$ (black).
    \emph{Bottom left panels:} weighted histogram of the best-fit progenitor masses.
    \emph{Bottom right panels:} the model energy and dilution masses. The best fit model is shown as a blue point. The dashed red line indicates the minimum dilution mass for a given energy. See text for details.}
    \label{fig:starfit}
\end{figure}

The two stars CarII-5664 and CarIII-1120 have low enough [Fe/H] that they are plausibly enriched \emph{only} by Pop~III stars \citep[e.g.,][]{Frebel15}.
Under this assumption, we fit models from \citet{Heger10} to the data to estimate the initial progenitor mass, explosion energy, internal mixing, and gas dilution mass for these stars.
To summarize the fitting procedure, we find the optimum dilution mass for all 16800 models in the \citet{Heger10} grid, reject all models inconsistent with our upper limits, then weight each remaining model by using its deviation from the best-fit $\chi^2$ as input to a $\chi^2$ survival function with 4 degrees of freedom.
The detailed fitting procedure and parameter description is described in \citet{Frebel19}\footnote{Code at \url{https://github.com/alexji/alexmods/blob/master/alexmods/alex_starfit.py}}.
Here, we exclude the elements Al, K, and Mn due to the uncertain size of NLTE corrections; and the elements Sc, Cr, Cu, and Zn due to model calculation uncertainties \citep{Heger10}.
Abundance corrections to C, Na, and Mg have been included (Table~\ref{tab:corr}).
We note that the \citet{Heger10} models do not include stellar rotation. However, rotation can substantially influence stellar evolution and the resulting nucleosynthesis \citep[e.g.,][]{Maeder15} and should be considered in future analyses.

The results are shown in Figure~\ref{fig:starfit}. We plot all models within $2\sigma$ contours of $\chi^2$ (i.e., models with weight $\gtrsim 0.05$).
In the top panel for each star, we show the data as filled red squares with error bars and upper limits as downward pointing arrows. Unused measurements and upper limits are indicated as open squares and downward pointing triangles, respectively.
The best-fit model is shown as a solid blue line, while other models within $2\sigma$ are shown as black lines.
For visualization purposes, models with worse $\chi^2$ are plotted as thinner transparent lines.
The bottom left panel for each star shows the weighted histogram for the resulting progenitor masses of the full fit.
The bottom right panel shows the best-fit energy and dilution masses, where again models with worse $\chi^2$ are displayed as smaller and more transparent points. The best fit model is again shown as a solid blue point.
In general, satisfactory fits were found for these two stars with $\mbox{[Fe/H]} < -3.5$.
CarIII-1120 is most consistent with a relatively low mass progenitor between $10-20 M_\odot$ with a typical ${\sim}1\times10^{51}$ erg explosion energy. Note that CarIII-1120 is a Group 2 CEMP-no star according to \citet{Yoon16}.
CarII-5664 is also best fit by a similar low-mass progenitor, but most of the best-fit models actually prefer a higher mass progenitor of $25-35 M_\odot$ with slightly higher explosion energy.

The combination of explosion energy and dilution mass introduces another consistency check.
A supernova with explosion energy $E$ will produce a supernova remnant that sweeps up a certain amount of mass before merging with the ISM \citep[e.g.,][]{Cioffi88}. This is the \emph{minimum} dilution mass allowable for that explosion energy (assuming no rare interactions such as colliding supernova blastwaves).
In the bottom right panels of Figure~\ref{fig:starfit} we show the approximate swept-up mass of a supernova remnant expanding into an efficiently cooling ISM $M_{\rm dil, H} = 0.75 \times 10^{4.5} M_\odot (E/10^{51}\,\text{erg})^{0.95}$ as a dotted red line \citep{Cioffi88, Ryan96}.
Models below this line are inconsistent with the explosion energy (though could be explained with enrichment by multiple supernovae), while models above the line are diluted beyond the supernova remnant due to turbulent mixing.
Applying this constraint tends to prefer higher explosion energies and higher masses.
In general, the best-fit dilution masses satisfying this constraint are ${\sim}10^5M_\odot$, suggesting that recollapsed gas within a minihalo is the most likely explanation for the origin of these stars rather than external pollution, as externally polluted halos have higher effective dilution masses \citep[e.g.,][]{Cooke14,Ji15,Smith15,Griffen18}.

\section{Conclusion}\label{sec:conclusion}

We present a comprehensive abundance analysis of the Magellanic satellite galaxies Carina~II and Carina~III using high-resolution Magellan/MIKE data, including the first abundances of an RR Lyrae star in any UFD.
The abundance results are shown in Figures~\ref{fig:grid} and \ref{fig:ncap}.
The stars in these two dwarf galaxies clearly do not show light element anticorrelations associated with globular clusters (Figure~\ref{fig:gcabund}).

The most notable chemical evolution trend is the variations in different $\alpha$-element ratios.
Car~II clearly shows different trends in [Mg/Fe] and [Ca/Fe] (Figure~\ref{fig:mgca}).
The origin of this evolution could be differences in core collapse and/or Type Ia supernova yields, and it is not yet clear which.
However, there are obvious differences in the [Mg/Ca] trends between different UFDs (Figure~\ref{fig:mgcaenv}), and we tentatively suggest this could be an environment-dependent abundance signature as LMC satellite UFDs have a different trend than MW satellite UFDs.
This suggestion will require studying the abundances of additional LMC satellites to confirm.

The most metal-poor stars in UFDs may contain signatures of the first metal-free Population~III stars.
Studying the whole population of Fe-poor UFD stars, we find that the carbon-enhanced fraction of UFD stars is essentially the same as the Milky Way halo (Figure~\ref{fig:cfrac}).
But, not all of the most Fe-poor stars in UFDs are carbon-enhanced: the most Fe-poor star in Car~II is clearly carbon-normal.
We also found two new stars with $\mbox{[Fe/H]} \leq -3.5$, bringing the total number of such stars in UFDs up to 8. The abundances of these stars are well-fit by Pop~III core-collapse supernova yields (Figure~\ref{fig:starfit}).

Our analysis of Car~II and III, along with the past decade of observations, brings the total number of UFD stars with high-resolution abundances up to ${\sim}85$ stars across $16$ different UFDs, of which now 5 UFDs have a ``large'' (${\geq}7$) number of stars studied (see references in Section~\ref{sec:abundsummary}).
While these data have already provided key insights into early nucleosynthesis and galaxy formation and pointed to many interesting abundance trends and signatures, the numbers of stars are still relatively small.
These sample sizes are currently dictated by the limits of current large telescopes, but 30m class telescopes will allow high-resolution spectroscopic abundances for 10s$-$100s of stars per UFD out to the virial radius of the Milky Way \citep{Ji19Decadal}, transforming our ability to unravel the detailed history of these first galaxy relics.

\acknowledgments
We thank Andy McWilliam, Evan Kirby, Ian Thompson, George Preston, Dan Kelson, Andrew Emerick, and Thomas Nordlander, and Chris Sneden for fruitful discussions; and Eduardo Ba\~nados for saving our RRL observations from certain doom.
APJ and TSL are supported by NASA through Hubble Fellowship grant HST-HF2-51393.001 and HST-HF2-51439.001 respectively, awarded by the Space Telescope Science Institute, which is operated by the Association of Universities for Research in Astronomy, Inc., for NASA, under contract NAS5-26555.
JDS is supported by the National Science Foundation under grant AST-1714873.
SK is partially supported by NSF awards AST-1813881 and AST-1909584.
The work of ABP is supported by NSF grant AST-1813881.
This research has made use of NASA's Astrophysics Data System Bibliographic Services.

This work has made use of data from the European Space Agency (ESA) mission
{\it Gaia} (\url{https://www.cosmos.esa.int/gaia}), processed by the {\it Gaia}
Data Processing and Analysis Consortium (DPAC,
\url{https://www.cosmos.esa.int/web/gaia/dpac/consortium}). Funding for the DPAC
has been provided by national institutions, in particular the institutions
participating in the {\it Gaia} Multilateral Agreement.

This project used data obtained with the Dark Energy Camera (DECam),
which was constructed by the Dark Energy Survey (DES) collaboration.
Funding for the DES Projects has been provided by 
the U.S. Department of Energy, 
the U.S. National Science Foundation, 
the Ministry of Science and Education of Spain, 
the Science and Technology Facilities Council of the United Kingdom, 
the Higher Education Funding Council for England, 
the National Center for Supercomputing Applications at the University of Illinois at Urbana-Champaign, 
the Kavli Institute of Cosmological Physics at the University of Chicago, 
the Center for Cosmology and Astro-Particle Physics at the Ohio State University, 
the Mitchell Institute for Fundamental Physics and Astronomy at Texas A\&M University, 
Financiadora de Estudos e Projetos, Funda{\c c}{\~a}o Carlos Chagas Filho de Amparo {\`a} Pesquisa do Estado do Rio de Janeiro, 
Conselho Nacional de Desenvolvimento Cient{\'i}fico e Tecnol{\'o}gico and the Minist{\'e}rio da Ci{\^e}ncia, Tecnologia e Inovac{\~a}o, 
the Deutsche Forschungsgemeinschaft, 
and the Collaborating Institutions in the Dark Energy Survey. 
The Collaborating Institutions are 
Argonne National Laboratory, 
the University of California at Santa Cruz, 
the University of Cambridge, 
Centro de Investigaciones En{\'e}rgeticas, Medioambientales y Tecnol{\'o}gicas-Madrid, 
the University of Chicago, 
University College London, 
the DES-Brazil Consortium, 
the University of Edinburgh, 
the Eidgen{\"o}ssische Technische Hoch\-schule (ETH) Z{\"u}rich, 
Fermi National Accelerator Laboratory, 
the University of Illinois at Urbana-Champaign, 
the Institut de Ci{\`e}ncies de l'Espai (IEEC/CSIC), 
the Institut de F{\'i}sica d'Altes Energies, 
Lawrence Berkeley National Laboratory, 
the Ludwig-Maximilians Universit{\"a}t M{\"u}nchen and the associated Excellence Cluster Universe, 
the University of Michigan, 
{the} National Optical Astronomy Observatory,
the University of Nottingham, 
the Ohio State University, 
the OzDES Membership Consortium
the University of Pennsylvania, 
the University of Portsmouth, 
SLAC National Accelerator Laboratory, 
Stanford University, 
the University of Sussex, 
and Texas A\&M University.

Based on observations at Cerro Tololo Inter-American Observatory, National Optical
Astronomy Observatory (NOAO Prop. ID 2016A-0366 and PI Keith Bechtol), which is operated by the Association of
Universities for Research in Astronomy (AURA) under a cooperative agreement with the
National Science Foundation.
This project is partially supported by the NASA Fermi Guest Investigator Program Cycle 9 No. 91201. 

\facilities{Magellan-Clay (MIKE)}
\software{MOOG \citep{Sneden73,Sobeck11}, SMHR \citep{Casey14}, \texttt{numpy} \citep{numpy}, 
\texttt{scipy} \citep{scipy}, 
\texttt{matplotlib} \citep{matplotlib},
\texttt{pandas} \citep{pandas},
\texttt{seaborn}, \citep{seaborn},
\texttt{astropy} \citep{astropy}}

\appendix

\section{Abundance Error Analysis Formalism}\label{app:error}

Here we explicitly list the equations used for our error analysis.
For element $X$, with lines indexed by $i$ that have abundances $A_i$, statistical error $\sigma_{i,\text{stat}}$, and systematic abundance offsets $\delta_{i,\Teff}$, $\delta_{i,\logg}$, $\delta_{i,\nu_t}$ and $\delta_{i,\text{[M/H]}}$ (note that the systematic abundance offsets retain their sign so we refer to them as $\delta_i$):
\begin{align}
    \sigma_{i,\text{sys}}^2 &= \delta_{i,\Teff}^2 + \delta_{i,\logg}^2 + \delta_{i,\nu_t}^2 + \delta_{i,\text{[M/H]}}^2 \\
    &\equiv \sum_{SP} \delta_{i,SP}^2 \\
    \sigma_i^2 &= \sigma_{i,\text{stat}}^2 + \sigma_{i,\text{sys}}^2 \label{eq:sigmai}
\end{align}
The statistical error $\sigma_{i, \text{stat}}$ quantifies the spectrum noise, either through the $1\sigma$ equivalent width uncertainty or $\chi^2$ uncertainty for synthesis.
Our equivalent width and synthesis fits allow the local continuum to vary by a linear function, using $\chi^2$ minimization to find the continuum level. Our quoted statistical uncertainties $\sigma_{i,\text{stat}}$ propagate these continuum uncertainties, and they match those inferred from simpler formulas based on the line FWHM within 5\% \citep[e.g.,][]{Battaglia08,Frebel06a}.

It is in principle possible that our local spectrum models are not accurate, and
the most impactful systematic would be misplacing the overall continuum level. As an extra conservative error bar, we include an additional column $\sigma_{\rm cont}$ in Table~\ref{tab:lines}, which is the uncertainty from systematically changing the overall continuum by the local 1$\sigma$ spectrum noise (i.e., the abundance difference after multiplying each equivalent width by $1 \pm 1/\text{SNR}$).
For synthesis measurements, we estimate this uncertainty by calculating the equivalent width of the synthetic feature without any other elements, then treating it as an equivalent width measurement. We thus did not estimate the continuum error for the molecular features.
A very conservative error estimate would also add this error in quadrature as part of equation~\ref{eq:sigmai}.
However, we are confident that our continuum placement procedure uncertainties are accurately reflected in the statistical error bar, so we do not include $\sigma_{\rm cont}$ in our abundance uncertainties.

We then assign each line a weight $w_i$
\begin{equation}
    w_i = \sigma_i^{-2}
\end{equation}
We adopt the weighted average of the lines as the final abundance, with statistical and systematic uncertainties:
\begin{align}
    A(X) &= \frac{\sum_i w_iA_i}{\sum_i w_i} \\
    \sigma_{\text{stat}}^2(X) &= \frac{\sum_i w_i (A_i-A(X))^2}{\sum_i w_i} + \frac{1}{\sum_i w_i} \\
    \delta_{\text{sys},SP}(X) &= \frac{\sum_i w_i (A_i + \delta_{i,SP})}{\sum_i w_i} - A(X) \\
                              &= \frac{\sum_i w_i \delta_{i,SP}}{\sum_i w_i}
\end{align}
The total statistical uncertainty accounts for both noise in individual lines as well as the weighted standard error of different lines.
Here we adopt just the first order Taylor expansion for the stellar parameter uncertainty, neglecting covariance between stellar parameters (see \citealt{McWilliam13}).
Finally, the total abundance error for [X/H] and element ratios [X/Y] combines the statistical and systematic uncertainties in quadrature:
\begin{align}
    \sigma_{\text{[X/H]}}^2 &= \sigma_{\text{stat}}^2 + \sum_{SP} \delta_{\text{sys},SP}^2 \\
    \sigma_{\text{[X/Y]}}^2 &= \sigma_{X,\text{stat}}^2 + \sigma_{Y,\text{stat}}^2 + \sum_{SP} \left(\delta_{X,SP} - \delta_{Y,SP}\right)^2
\end{align}
Note that for an element ratio of X and Y, we only allow covariance between $X$ and $Y$ through the stellar parameters.

\section{Abundance Tables}\label{app:abunds}
\startlongtable
\begin{deluxetable}{lccrrrrrr}
\tablecolumns{9}
\tabletypesize{\scriptsize}
\tablecaption{\label{tab:abunds}Stellar Abundances}
\tablehead{El. & $N$ &  & $\log \epsilon$ & $\sigma_{\text{stat}}$ & [X/H] & $\sigma_{\text{[X/H]}}$ & [X/Fe] & $\sigma_{\text{[X/Fe]}}$}
\startdata
\cutinhead{CarII-0064}
Na I & 2 &     & 3.65 &  0.15 & -2.59 &  0.31 & -0.36 &  0.19 \\
Mg I & 5 &     & 5.21 &  0.07 & -2.39 &  0.17 & -0.17 &  0.08 \\
Al I & 2 &     & 3.14 &  0.23 & -3.31 &  0.30 & -1.09 &  0.30 \\
Si I & 2 &     & 5.19 &  0.27 & -2.32 &  0.36 & -0.10 &  0.28 \\
Ca I & 22 &     & 4.31 &  0.03 & -2.03 &  0.14 & 0.20 &  0.06 \\
Sc II & 6 &     & 0.00 &  0.04 & -3.15 &  0.13 & -0.94 &  0.07 \\
Ti I & 16 &     & 2.54 &  0.05 & -2.41 &  0.24 & -0.19 &  0.08 \\
Ti II & 36 &     & 2.67 &  0.02 & -2.29 &  0.13 & -0.08 &  0.07 \\
Cr I & 15 &     & 3.31 &  0.04 & -2.33 &  0.22 & -0.11 &  0.06 \\
Cr II & 2 &     & 3.66 &  0.07 & -1.98 &  0.14 & 0.22 &  0.09 \\
Mn I & 6 &     & 2.53 &  0.06 & -2.90 &  0.16 & -0.68 &  0.07 \\
Fe I & 169 &     & 5.28 &  0.01 & -2.22 &  0.18 & 0.00 &  0.02 \\
Fe II & 21 &     & 5.30 &  0.04 & -2.20 &  0.13 & 0.00 &  0.05 \\
Co I & 5 &     & 2.60 &  0.11 & -2.39 &  0.22 & -0.16 &  0.11 \\
Ni I & 8 &     & 3.75 &  0.04 & -2.47 &  0.16 & -0.25 &  0.06 \\
Zn I & 2 &     & 2.21 &  0.14 & -2.35 &  0.16 & -0.13 &  0.20 \\
Sr II & 2 &     & -0.71 &  0.23 & -3.58 &  0.39 & -1.38 &  0.32 \\
Ba II & 2 &     & -2.58 &  0.13 & -4.76 &  0.21 & -2.55 &  0.20 \\
C-H & 2 &     & 6.16 &  0.16 & -2.27 &  0.35 & -0.05 &  0.22 \\
C-N & 1 &     & 6.04 &  0.65 & -1.79 &  0.78 & 0.43 &  0.70 \\
O I & 1 & $<$ & 7.59 &\nodata& -1.09 &\nodata& 1.13 &\nodata\\
K I & 1 & $<$ & 2.75 &\nodata& -2.28 &\nodata& -0.06 &\nodata\\
Cu I & 1 & $<$ & 1.83 &\nodata& -2.36 &\nodata& -0.14 &\nodata\\
Eu II & 1 & $<$ & -1.89 &\nodata& -2.41 &\nodata& -0.21 &\nodata\\
\cutinhead{CarII-2064}
Na I & 2 &     & 4.21 &  0.15 & -2.03 &  0.33 & 0.33 &  0.20 \\
Mg I & 5 &     & 5.62 &  0.12 & -1.98 &  0.22 & 0.38 &  0.13 \\
Al I & 2 &     & 3.37 &  0.40 & -3.08 &  0.44 & -0.73 &  0.41 \\
Si I & 2 &     & 6.27 &  0.48 & -1.24 &  0.58 & 1.11 &  0.51 \\
K I & 1 &     & 3.46 &  0.18 & -1.57 &  0.23 & 0.79 &  0.18 \\
Ca I & 14 &     & 4.48 &  0.06 & -1.86 &  0.15 & 0.49 &  0.09 \\
Sc II & 6 &     & 0.96 &  0.12 & -2.19 &  0.19 & 0.12 &  0.13 \\
Ti I & 11 &     & 3.43 &  0.05 & -1.52 &  0.23 & 0.83 &  0.06 \\
Ti II & 26 &     & 3.23 &  0.05 & -1.72 &  0.17 & 0.59 &  0.10 \\
Cr I & 3 &     & 2.73 &  0.35 & -2.91 &  0.44 & -0.55 &  0.36 \\
Cr II & 1 &     & 3.78 &  0.18 & -1.86 &  0.23 & 0.44 &  0.20 \\
Mn I & 3 &     & 3.02 &  0.16 & -2.42 &  0.26 & -0.06 &  0.17 \\
Fe I & 81 &     & 5.15 &  0.03 & -2.35 &  0.20 & 0.00 &  0.04 \\
Fe II & 10 &     & 5.19 &  0.06 & -2.31 &  0.15 & 0.00 &  0.09 \\
Co I & 1 &     & 2.43 &  0.28 & -2.56 &  0.35 & -0.20 &  0.29 \\
Ni I & 2 &     & 4.31 &  0.17 & -1.91 &  0.24 & 0.44 &  0.17 \\
Sr II & 2 &     & -1.08 &  0.20 & -3.95 &  0.26 & -1.64 &  0.22 \\
Ba II & 2 &     & -0.97 &  0.11 & -3.15 &  0.20 & -0.84 &  0.16 \\
C-H & 2 &     & 6.45 &  0.21 & -1.98 &  0.45 & 0.38 &  0.31 \\
O I & 2 & $<$ & 8.73 &\nodata& 0.04 &\nodata& 2.40 &\nodata\\
Cu I & 2 & $<$ & 4.10 &\nodata& -0.09 &\nodata& 2.27 &\nodata\\
Zn I & 2 & $<$ & 2.76 &\nodata& -1.80 &\nodata& 0.55 &\nodata\\
Eu II & 2 & $<$ & -0.64 &\nodata& -1.16 &\nodata& 1.15 &\nodata\\
C-N & 2 & $<$ & 4.48 &\nodata& -3.35 &\nodata& -1.00 &\nodata\\
\cutinhead{CarII-4704}
Na I & 2 &     & 3.50 &  0.16 & -2.74 &  0.33 & -0.54 &  0.19 \\
Mg I & 5 &     & 5.06 &  0.12 & -2.54 &  0.19 & -0.34 &  0.14 \\
Al I & 2 &     & 3.00 &  0.56 & -3.45 &  0.58 & -1.25 &  0.58 \\
Si I & 2 &     & 4.96 &  0.41 & -2.55 &  0.47 & -0.35 &  0.41 \\
K I & 1 &     & 2.96 &  0.17 & -2.07 &  0.23 & 0.13 &  0.18 \\
Ca I & 14 &     & 4.20 &  0.04 & -2.14 &  0.15 & 0.06 &  0.07 \\
Sc II & 5 &     & 0.74 &  0.08 & -2.41 &  0.15 & -0.22 &  0.10 \\
Ti I & 5 &     & 2.65 &  0.09 & -2.30 &  0.24 & -0.10 &  0.10 \\
Ti II & 26 &     & 2.65 &  0.05 & -2.30 &  0.14 & -0.12 &  0.09 \\
Cr I & 9 &     & 3.23 &  0.09 & -2.41 &  0.23 & -0.20 &  0.09 \\
Mn I & 7 &     & 2.58 &  0.09 & -2.85 &  0.18 & -0.64 &  0.10 \\
Fe I & 105 &     & 5.30 &  0.02 & -2.20 &  0.20 & 0.00 &  0.03 \\
Fe II & 14 &     & 5.32 &  0.04 & -2.19 &  0.12 & 0.00 &  0.05 \\
Co I & 4 &     & 2.69 &  0.15 & -2.30 &  0.24 & -0.10 &  0.15 \\
Ni I & 3 &     & 4.11 &  0.09 & -2.11 &  0.18 & 0.09 &  0.10 \\
Sr II & 2 &     & -1.68 &  0.26 & -4.55 &  0.31 & -2.36 &  0.27 \\
Ba II & 2 &     & -2.11 &  0.11 & -4.29 &  0.18 & -2.10 &  0.17 \\
C-H & 2 &     & 5.63 &  0.16 & -2.80 &  0.35 & -0.60 &  0.22 \\
O I & 2 & $<$ & 7.66 &\nodata& -1.03 &\nodata& 1.17 &\nodata\\
Cu I & 2 & $<$ & 2.69 &\nodata& -1.50 &\nodata& 0.70 &\nodata\\
Zn I & 2 & $<$ & 2.88 &\nodata& -1.69 &\nodata& 0.52 &\nodata\\
Eu II & 2 & $<$ & -1.44 &\nodata& -1.96 &\nodata& 0.23 &\nodata\\
C-N & 2 & $<$ & 5.75 &\nodata& -2.08 &\nodata& 0.13 &\nodata\\
\cutinhead{CarII-4928}
Na I & 2 &     & 3.45 &  0.16 & -2.79 &  0.35 & 0.26 &  0.20 \\
Mg I & 5 &     & 5.25 &  0.12 & -2.35 &  0.21 & 0.71 &  0.16 \\
Al I & 2 &     & 3.46 &  0.35 & -2.99 &  0.45 & 0.07 &  0.40 \\
Si I & 2 &     & 4.81 &  0.31 & -2.70 &  0.40 & 0.35 &  0.31 \\
Ca I & 2 &     & 4.08 &  0.18 & -2.26 &  0.28 & 0.80 &  0.20 \\
Sc II & 6 &     & 0.36 &  0.11 & -2.79 &  0.23 & 0.24 &  0.19 \\
Ti II & 22 &     & 2.58 &  0.09 & -2.37 &  0.22 & 0.66 &  0.18 \\
Cr I & 3 &     & 1.90 &  0.20 & -3.74 &  0.37 & -0.68 &  0.21 \\
Mn I & 3 &     & 1.47 &  0.16 & -3.96 &  0.30 & -0.91 &  0.17 \\
Fe I & 46 &     & 4.44 &  0.04 & -3.06 &  0.26 & 0.00 &  0.05 \\
Fe II & 5 &     & 4.47 &  0.14 & -3.03 &  0.23 & 0.00 &  0.20 \\
Co I & 2 &     & 2.09 &  0.24 & -2.90 &  0.39 & 0.16 &  0.26 \\
Sr II & 2 &     & -1.99 &  0.25 & -4.86 &  0.33 & -1.83 &  0.30 \\
C-H & 2 &     & 5.67 &  0.26 & -2.77 &  0.56 & 0.29 &  0.39 \\
O I & 2 & $<$ & 8.50 &\nodata& -0.19 &\nodata& 2.86 &\nodata\\
K I & 2 & $<$ & 3.04 &\nodata& -1.99 &\nodata& 1.06 &\nodata\\
Ni I & 2 & $<$ & 5.16 &\nodata& -1.06 &\nodata& 1.99 &\nodata\\
Cu I & 2 & $<$ & 3.06 &\nodata& -1.13 &\nodata& 1.93 &\nodata\\
Zn I & 2 & $<$ & 3.08 &\nodata& -1.48 &\nodata& 1.58 &\nodata\\
Ba II & 2 & $<$ & -1.60 &\nodata& -3.78 &\nodata& -0.74 &\nodata\\
Eu II & 2 & $<$ & -1.04 &\nodata& -1.56 &\nodata& 1.47 &\nodata\\
C-N & 2 & $<$ & 6.97 &\nodata& -0.86 &\nodata& 2.19 &\nodata\\
\cutinhead{CarII-5664}
Na I & 2 &     & 3.37 &  0.12 & -2.87 &  0.27 & 0.66 &  0.17 \\
Mg I & 7 &     & 4.80 &  0.04 & -2.79 &  0.12 & 0.73 &  0.09 \\
Al I & 2 &     & 2.46 &  0.17 & -3.99 &  0.29 & -0.46 &  0.27 \\
Si I & 2 &     & 4.59 &  0.13 & -2.92 &  0.26 & 0.61 &  0.14 \\
K I & 2 &     & 2.52 &  0.07 & -2.51 &  0.16 & 1.02 &  0.09 \\
Ca I & 12 &     & 3.18 &  0.03 & -3.16 &  0.13 & 0.36 &  0.07 \\
Sc II & 8 &     & -0.14 &  0.06 & -3.29 &  0.12 & 0.24 &  0.08 \\
Ti I & 6 &     & 1.64 &  0.04 & -3.31 &  0.22 & 0.21 &  0.07 \\
Ti II & 32 &     & 1.68 &  0.03 & -3.27 &  0.11 & 0.27 &  0.07 \\
Cr I & 3 &     & 1.47 &  0.09 & -4.17 &  0.25 & -0.64 &  0.12 \\
Mn I & 3 &     & 0.91 &  0.11 & -4.52 &  0.25 & -0.99 &  0.12 \\
Fe I & 113 &     & 3.97 &  0.01 & -3.53 &  0.18 & 0.00 &  0.02 \\
Fe II & 10 &     & 3.96 &  0.04 & -3.54 &  0.10 & 0.00 &  0.05 \\
Co I & 4 &     & 1.47 &  0.07 & -3.52 &  0.22 & 0.00 &  0.08 \\
Ni I & 1 &     & 2.25 &  0.20 & -3.96 &  0.28 & -0.44 &  0.21 \\
Sr II & 2 &     & -2.86 &  0.09 & -5.73 &  0.16 & -2.19 &  0.12 \\
C-H & 2 &     & 4.74 &  0.21 & -3.69 &  0.48 & -0.17 &  0.33 \\
O I & 2 & $<$ & 6.88 &\nodata& -1.81 &\nodata& 1.72 &\nodata\\
Cu I & 2 & $<$ & 1.47 &\nodata& -2.72 &\nodata& 0.80 &\nodata\\
Zn I & 2 & $<$ & 1.54 &\nodata& -3.02 &\nodata& 0.50 &\nodata\\
Ba II & 2 & $<$ & -3.29 &\nodata& -5.47 &\nodata& -1.93 &\nodata\\
Eu II & 2 & $<$ & -2.81 &\nodata& -3.33 &\nodata& 0.21 &\nodata\\
C-N & 2 & $<$ & 5.79 &\nodata& -2.04 &\nodata& 1.48 &\nodata\\
\cutinhead{CarII-6544}
O I & 2 &     & 6.99 &  0.09 & -1.70 &  0.18 & 0.96 &  0.21 \\
Na I & 2 &     & 3.53 &  0.20 & -2.71 &  0.44 & -0.05 &  0.28 \\
Mg I & 7 &     & 5.11 &  0.08 & -2.49 &  0.16 & 0.17 &  0.10 \\
Al I & 2 &     & 3.17 &  0.25 & -3.28 &  0.36 & -0.62 &  0.26 \\
Si I & 2 &     & 5.16 &  0.25 & -2.35 &  0.36 & 0.31 &  0.26 \\
K I & 1 &     & 2.69 &  0.20 & -2.34 &  0.28 & 0.32 &  0.20 \\
Ca I & 21 &     & 3.78 &  0.03 & -2.56 &  0.15 & 0.10 &  0.07 \\
Sc II & 11 &     & 0.33 &  0.04 & -2.82 &  0.11 & -0.16 &  0.06 \\
Ti I & 22 &     & 2.19 &  0.04 & -2.77 &  0.29 & -0.11 &  0.10 \\
Ti II & 44 &     & 2.34 &  0.02 & -2.61 &  0.12 & 0.05 &  0.08 \\
Cr I & 16 &     & 2.84 &  0.03 & -2.80 &  0.27 & -0.15 &  0.07 \\
Cr II & 1 &     & 3.12 &  0.14 & -2.52 &  0.19 & 0.14 &  0.15 \\
Mn I & 7 &     & 2.29 &  0.06 & -3.14 &  0.22 & -0.48 &  0.07 \\
Fe I & 144 &     & 4.84 &  0.01 & -2.66 &  0.21 & 0.00 &  0.02 \\
Fe II & 22 &     & 4.84 &  0.03 & -2.66 &  0.13 & 0.00 &  0.05 \\
Co I & 5 &     & 2.19 &  0.12 & -2.80 &  0.27 & -0.14 &  0.13 \\
Ni I & 12 &     & 3.43 &  0.04 & -2.79 &  0.20 & -0.13 &  0.05 \\
Zn I & 1 &     & 1.47 &  0.13 & -3.09 &  0.15 & -0.43 &  0.23 \\
Sr II & 2 &     & -1.58 &  0.15 & -4.45 &  0.25 & -1.79 &  0.21 \\
Ba II & 4 &     & -2.00 &  0.10 & -4.18 &  0.16 & -1.52 &  0.15 \\
C-H & 2 &     & 5.24 &  0.16 & -3.19 &  0.35 & -0.54 &  0.21 \\
C-N & 1 &     & 5.48 &  0.54 & -2.35 &  0.67 & 0.31 &  0.58 \\
Cu I & 1 & $<$ & 1.48 &\nodata& -2.71 &\nodata& -0.06 &\nodata\\
Eu II & 1 & $<$ & -2.34 &\nodata& -2.86 &\nodata& -0.20 &\nodata\\
\cutinhead{CarII-7872}
O I & 2 &     & 7.56 &  0.07 & -1.13 &  0.15 & 1.38 &  0.21 \\
Na I & 2 &     & 3.21 &  0.18 & -3.03 &  0.39 & -0.52 &  0.23 \\
Mg I & 5 &     & 5.09 &  0.06 & -2.51 &  0.20 & -0.00 &  0.08 \\
Al I & 2 &     & 3.59 &  0.27 & -2.86 &  0.43 & -0.35 &  0.31 \\
Si I & 2 &     & 5.17 &  0.32 & -2.34 &  0.43 & 0.17 &  0.34 \\
K I & 2 &     & 2.83 &  0.12 & -2.20 &  0.25 & 0.31 &  0.12 \\
Ca I & 12 &     & 3.83 &  0.04 & -2.51 &  0.17 & 0.00 &  0.07 \\
Sc II & 5 &     & 0.54 &  0.06 & -2.61 &  0.16 & -0.12 &  0.08 \\
Ti I & 18 &     & 2.38 &  0.04 & -2.57 &  0.30 & -0.06 &  0.10 \\
Ti II & 27 &     & 2.55 &  0.03 & -2.40 &  0.12 & 0.09 &  0.09 \\
Cr I & 16 &     & 2.86 &  0.04 & -2.78 &  0.27 & -0.27 &  0.07 \\
Mn I & 6 &     & 2.29 &  0.08 & -3.13 &  0.20 & -0.62 &  0.10 \\
Fe I & 123 &     & 4.99 &  0.02 & -2.51 &  0.21 & 0.00 &  0.02 \\
Fe II & 20 &     & 5.01 &  0.04 & -2.49 &  0.13 & 0.00 &  0.06 \\
Co I & 5 &     & 2.36 &  0.10 & -2.63 &  0.22 & -0.12 &  0.11 \\
Ni I & 10 &     & 3.62 &  0.05 & -2.60 &  0.20 & -0.08 &  0.06 \\
Zn I & 1 &     & 1.73 &  0.10 & -2.83 &  0.12 & -0.32 &  0.23 \\
Sr II & 1 &     & -1.03 &  0.36 & -3.90 &  0.46 & -1.41 &  0.41 \\
Ba II & 5 &     & -0.89 &  0.06 & -3.07 &  0.16 & -0.57 &  0.13 \\
C-H & 2 &     & 6.24 &  0.14 & -2.19 &  0.31 & 0.32 &  0.20 \\
C-N & 1 &     & 6.48 &  0.53 & -1.35 &  0.55 & 1.17 &  0.55 \\
Cu I & 1 & $<$ & 1.42 &\nodata& -2.77 &\nodata& -0.25 &\nodata\\
Eu II & 1 & $<$ & -2.59 &\nodata& -3.11 &\nodata& -0.61 &\nodata\\
\cutinhead{CarII-9296}
Na I & 2 &     & 3.38 &  0.15 & -2.86 &  0.33 & 0.03 &  0.19 \\
Mg I & 5 &     & 5.18 &  0.09 & -2.42 &  0.21 & 0.47 &  0.11 \\
Al I & 2 &     & 3.42 &  0.35 & -3.04 &  0.44 & -0.15 &  0.40 \\
Si I & 2 &     & 4.71 &  0.28 & -2.80 &  0.39 & 0.09 &  0.29 \\
Ca I & 12 &     & 3.82 &  0.05 & -2.52 &  0.17 & 0.37 &  0.09 \\
Sc II & 6 &     & 0.37 &  0.07 & -2.78 &  0.18 & 0.13 &  0.11 \\
Ti I & 4 &     & 2.60 &  0.15 & -2.35 &  0.30 & 0.54 &  0.16 \\
Ti II & 18 &     & 2.49 &  0.05 & -2.46 &  0.18 & 0.45 &  0.12 \\
Cr I & 4 &     & 2.28 &  0.15 & -3.36 &  0.35 & -0.47 &  0.18 \\
Mn I & 3 &     & 2.22 &  0.36 & -3.21 &  0.42 & -0.32 &  0.36 \\
Fe I & 72 &     & 4.61 &  0.03 & -2.89 &  0.23 & 0.00 &  0.05 \\
Fe II & 8 &     & 4.59 &  0.08 & -2.91 &  0.17 & 0.00 &  0.11 \\
Co I & 2 &     & 1.76 &  0.21 & -3.23 &  0.36 & -0.34 &  0.23 \\
Ni I & 1 &     & 3.24 &  0.34 & -2.98 &  0.42 & -0.09 &  0.34 \\
Sr II & 2 &     & -1.28 &  0.26 & -4.15 &  0.39 & -1.25 &  0.32 \\
Ba II & 3 &     & -1.95 &  0.14 & -4.13 &  0.25 & -1.23 &  0.22 \\
C-H & 2 &     & 5.72 &  0.32 & -2.71 &  0.56 & 0.17 &  0.41 \\
O I & 2 & $<$ & 7.88 &\nodata& -0.81 &\nodata& 2.08 &\nodata\\
K I & 2 & $<$ & 3.13 &\nodata& -1.90 &\nodata& 0.99 &\nodata\\
Cu I & 2 & $<$ & 2.38 &\nodata& -1.81 &\nodata& 1.07 &\nodata\\
Zn I & 2 & $<$ & 2.70 &\nodata& -1.86 &\nodata& 1.03 &\nodata\\
Eu II & 2 & $<$ & -1.44 &\nodata& -1.96 &\nodata& 0.95 &\nodata\\
C-N & 2 & $<$ & 6.31 &\nodata& -1.52 &\nodata& 1.37 &\nodata\\
\cutinhead{CarII-V3}
Na I & 2 &     & 3.04 &  0.13 & -3.20 &  0.27 & -0.56 &  0.15 \\
Mg I & 2 &     & 5.62 &  0.31 & -1.98 &  0.40 & 0.67 &  0.33 \\
Al I & 2 &     & 3.24 &  0.16 & -3.21 &  0.32 & -0.56 &  0.17 \\
Si I & 1 &     & 4.29 &  0.31 & -3.22 &  0.41 & -0.57 &  0.31 \\
Ca I & 1 &     & 4.01 &  0.36 & -2.33 &  0.49 & 0.31 &  0.38 \\
Sc II & 2 &     & 0.66 &  0.17 & -2.49 &  0.26 & 0.23 &  0.21 \\
Ti II & 14 &     & 2.53 &  0.05 & -2.42 &  0.18 & 0.30 &  0.11 \\
Cr I & 3 &     & 2.88 &  0.13 & -2.76 &  0.35 & -0.12 &  0.14 \\
Fe I & 22 &     & 4.85 &  0.04 & -2.65 &  0.28 & 0.00 &  0.06 \\
Fe II & 8 &     & 4.78 &  0.06 & -2.72 &  0.13 & 0.00 &  0.09 \\
\cutinhead{CarIII-1120}
Na I & 2 &     & 3.75 &  0.20 & -2.49 &  0.45 & 1.39 &  0.26 \\
Mg I & 5 &     & 4.94 &  0.10 & -2.66 &  0.23 & 1.22 &  0.13 \\
Al I & 2 &     & 2.41 &  0.37 & -4.04 &  0.46 & -0.17 &  0.45 \\
Si I & 1 &     & 4.72 &  0.42 & -2.79 &  0.50 & 1.08 &  0.42 \\
Ca I & 1 &     & 2.93 &  0.22 & -3.41 &  0.28 & 0.46 &  0.24 \\
Sc II & 5 &     & -0.60 &  0.12 & -3.75 &  0.18 & 0.14 &  0.16 \\
Ti II & 6 &     & 1.36 &  0.07 & -3.59 &  0.18 & 0.30 &  0.14 \\
Cr I & 2 &     & 1.29 &  0.17 & -4.34 &  0.33 & -0.47 &  0.18 \\
Mn I & 3 &     & 0.86 &  0.18 & -4.57 &  0.25 & -0.70 &  0.22 \\
Fe I & 48 &     & 3.63 &  0.03 & -3.87 &  0.27 & 0.00 &  0.05 \\
Fe II & 4 &     & 3.61 &  0.08 & -3.89 &  0.14 & 0.00 &  0.12 \\
Co I & 3 &     & 1.62 &  0.17 & -3.37 &  0.28 & 0.51 &  0.20 \\
Sr II & 2 &     & -2.75 &  0.24 & -5.62 &  0.28 & -1.72 &  0.26 \\
C-H & 2 &     & 5.81 &  0.28 & -2.62 &  0.63 & 1.25 &  0.43 \\
O I & 2 & $<$ & 7.50 &\nodata& -1.19 &\nodata& 2.68 &\nodata\\
K I & 2 & $<$ & 2.31 &\nodata& -2.72 &\nodata& 1.15 &\nodata\\
Ni I & 2 & $<$ & 3.12 &\nodata& -3.10 &\nodata& 0.78 &\nodata\\
Cu I & 2 & $<$ & 2.07 &\nodata& -2.12 &\nodata& 1.75 &\nodata\\
Zn I & 2 & $<$ & 2.55 &\nodata& -2.01 &\nodata& 1.86 &\nodata\\
Ba II & 2 & $<$ & -2.48 &\nodata& -4.66 &\nodata& -0.76 &\nodata\\
Eu II & 2 & $<$ & -1.82 &\nodata& -2.34 &\nodata& 1.55 &\nodata\\
C-N & 2 & $<$ & 5.15 &\nodata& -2.68 &\nodata& 1.20 &\nodata\\
\cutinhead{CarIII-8144}
Na I & 2 &     & 3.97 &  0.14 & -2.27 &  0.31 & -0.00 &  0.19 \\
Mg I & 7 &     & 5.58 &  0.08 & -2.02 &  0.18 & 0.25 &  0.10 \\
Al I & 2 &     & 3.58 &  0.23 & -2.87 &  0.27 & -0.60 &  0.24 \\
Si I & 2 &     & 5.84 &  0.18 & -1.67 &  0.32 & 0.60 &  0.21 \\
K I & 1 &     & 3.86 &  0.21 & -1.17 &  0.29 & 1.09 &  0.22 \\
Ca I & 24 &     & 4.62 &  0.03 & -1.72 &  0.14 & 0.55 &  0.05 \\
Sc II & 5 &     & 1.42 &  0.07 & -1.73 &  0.17 & 0.50 &  0.10 \\
Ti I & 15 &     & 3.05 &  0.02 & -1.90 &  0.21 & 0.36 &  0.05 \\
Ti II & 38 &     & 3.14 &  0.03 & -1.81 &  0.14 & 0.43 &  0.08 \\
Cr I & 13 &     & 3.23 &  0.05 & -2.41 &  0.20 & -0.14 &  0.06 \\
Cr II & 2 &     & 3.67 &  0.08 & -1.97 &  0.14 & 0.26 &  0.10 \\
Mn I & 7 &     & 2.97 &  0.11 & -2.46 &  0.19 & -0.19 &  0.12 \\
Fe I & 147 &     & 5.23 &  0.01 & -2.27 &  0.18 & 0.00 &  0.02 \\
Fe II & 16 &     & 5.27 &  0.04 & -2.23 &  0.13 & 0.00 &  0.06 \\
Co I & 3 &     & 2.56 &  0.14 & -2.43 &  0.24 & -0.16 &  0.16 \\
Ni I & 4 &     & 3.85 &  0.08 & -2.37 &  0.19 & -0.10 &  0.09 \\
Zn I & 2 &     & 2.84 &  0.08 & -1.72 &  0.12 & 0.55 &  0.15 \\
Sr II & 2 &     & -0.75 &  0.26 & -3.62 &  0.38 & -1.39 &  0.32 \\
Ba II & 1 &     & -2.07 &  0.14 & -4.25 &  0.18 & -2.02 &  0.17 \\
C-H & 2 &     & 6.31 &  0.16 & -2.12 &  0.35 & 0.15 &  0.23 \\
O I & 2 & $<$ & 8.17 &\nodata& -0.52 &\nodata& 1.74 &\nodata\\
Cu I & 2 & $<$ & 2.50 &\nodata& -1.69 &\nodata& 0.58 &\nodata\\
Eu II & 2 & $<$ & -1.30 &\nodata& -1.82 &\nodata& 0.42 &\nodata\\
C-N & 2 & $<$ & 6.03 &\nodata& -1.80 &\nodata& 0.46 &\nodata\\
\enddata
\end{deluxetable}

\ 
\end{document}